\documentclass[twocolumn]{aastex701}

\usepackage{amsmath}
\usepackage{mathtools}
\usepackage{tabularx}
\usepackage[hyphens]{url}
\usepackage{graphicx}
\usepackage{subfigure}
\usepackage[figuresright]{rotating}
\usepackage{booktabs}
\usepackage{multirow}

\received{XXX}
\revised{XXX}
\accepted{XXX}

\submitjournal{ApJ}

\shorttitle{MaDCoWS Splashback}
\shortauthors{Thongkham et al.}

\graphicspath{{./}{figures/}}

\begin{document}

\title{The Massive and Distant Clusters of WISE Survey 2: Splashback Radii to z=1.65 from Galaxy Density Profiles}

\author[0000-0001-7027-2202]{Khunanon Thongkham}
\email{khunanon@narit.or.th}
\affiliation{National Astronomical Research Institute of Thailand (NARIT), Mae Rim, Chiang Mai 50180, Thailand}
\affiliation{Korea Astronomy and Space Science Institute, 776, Daedeokdae-ro, Yuseong-gu, Daejeon 34055, Republic of Korea}
\affiliation{Department of Astronomy, University of Florida, 211 Bryant Space Center, Gainesville, FL 32611, USA}

\author[0000-0002-0933-8601]{Anthony H. Gonzalez}
\email{anthonyhg@astro.ufl.edu}
\affiliation{Department of Astronomy, University of Florida, 211 Bryant Space Center, Gainesville, FL 32611, USA}
 
\author[0000-0002-4208-798X]{Mark Brodwin}
\email{brodwin@eurekasci.com}
\affiliation{Eureka Scientific, 2452 Delmer Street Suite 100, Oakland, CA, 94602-3017, USA}

\author[0000-0003-3428-1106]{Ariane Trudeau}
\email{atrudeau@asiaa.sinica.edu.tw}
\affiliation{Department of Astronomy, University of Florida, 211 Bryant Space Center, Gainesville, FL 32611, USA}
\affiliation{Institute of Astronomy and Astrophysics, Academia Sinica, Taipei 10617, Taiwan}

\author{Peter Eisenhardt}
\email{Peter.R.Eisenhardt@jpl.nasa.gov}	
\affiliation{Jet Propulsion Laboratory, California Institute of Technology, 4800 Oak Grove Dr., Pasadena, CA 91109, USA}

\author[0000-0003-0122-0841]{S. A.\ Stanford}
\email{stanford@physics.ucdavis.edu}
\affiliation{Department of Physics, University of California, One Shields Avenue, Davis, CA 95616, USA}

\author[0000-0001-9793-5416]{Emily Moravec}
\email{emoravec@nrao.edu}
\affiliation{Green Bank Observatory, P.O. Box 2, Green Bank, WV 24944}

\author[0000-0002-7898-7664]{Thomas Connor}
\email{thomas.connor@cfa.harvard.edu}
\affiliation{Center for Astrophysics $\vert$\ Harvard\ \&\ Smithsonian, 60 Garden St., Cambridge, MA 02138, USA}

\author[0000-0003-2686-9241]{Daniel Stern}
\email{daniel.k.stern@jpl.nasa.gov}
\affiliation{Jet Propulsion Laboratory, California Institute of Technology, 4800 Oak Grove Dr., Pasadena, CA 91109, USA}

\begin{abstract}
The Massive and Distant Clusters of WISE Survey 2 (MaDCoWS2) is a WISE-selected catalog of galaxy clusters at $0.1<z<2$ covering an effective area of $>6000$ deg$^2$. In this paper, we derive splashback radii for this cluster ensemble from galaxy density profiles and constrain the mass threshold of the survey as a function of redshift. We use MaDCoWS2 cluster candidates at $0.4\leq z \leq 1.65$ divided into subsamples with different signal-to-noise (S/N$_{\rm P}$) and redshifts, cross-correlated with galaxies from the CatWISE2020 catalog, to obtain average surface density profiles. We perform a Markov Chain Monte Carlo analysis to derive parameter estimates for theoretical models consisting of orbiting and infalling terms. A distinct splashback feature is detected in all subsamples. The measured splashback radii span from $0.89^{+0.02}_{-0.02}h^{-1}$ comoving Mpc/cMpc ($0.61^{+0.02}_{-0.02}h^{-1}$ proper Mpc/pMpc) at $\overline{z}=0.45$ to $1.27^{+0.05}_{-0.05}h^{-1}$ cMpc ($0.53^{+0.04}_{-0.04}h^{-1}$ pMpc) at $\overline{z}=1.54$. We also find that splashback radii increase with $S/N_{\rm P}$ at fixed redshift. The resultant splashback radii constrain the redshift dependence of the mass of MaDCoWS2 clusters at fixed $S/N_{\rm P}$. We calculate $M_{\rm 200m}$ from the radii using a relation based on a cosmological simulation. MaDCoWS2 $M_{\rm 200m}$ values derived from the simulation-based relation are lower than the expected values based on weak-lensing observations. More robust mass constraints will come from calibrating splashback radii derived from galaxy density profiles with weak lensing shear profiles from facilities such as $\textit{Euclid}$, Rubin, and $\textit{Roman}$.
\end{abstract}


\section{Introduction} \label{sec:Introduction}
In recent decades, galaxy clusters have played a crucial role in advancing our understanding of cosmology \citep[e.g.][]{2009Vikhlinin,2011Allen} and astrophysics \citep{2010von,2012Fabian,2012McNamara,2022Alberts,2024Harrison}. Because gravitational collapse occurs more rapidly in a high-density universe, the number of galaxy clusters as a function of mass is a strong function of the matter density of the Universe. The amplitude of the mass function is also sensitive to the amplitude of the power spectrum on the scale of $8 h^{-1}$ Mpc or $\sigma_8$, which is related to matter density fluctuations \citep{2019Bocquet,2024Grandis}. The evolution of the mass function depends on the age of the Universe at a given redshift, and thus the mass function also probes dark energy \citep{2018Heneka,2018Pacaud}. Constraints derived from the mass function rely on having accurate, low-scatter cluster mass proxies because of the power law dependence of the mass function on cluster mass \citep{2008Tinker,2013Watson,2016Despali,2021Seppi}. A common approach to evaluating mass proxies is to define cluster masses and radii in terms of overdensity thresholds relative to the mean or critical density of the Universe. However, these mass definitions are not directly observable quantities. Furthermore, being tied to cosmology, they undergo pseudo-evolution as the matter density of the Universe changes with redshift. 

The splashback radius has been suggested by \cite{2014Diemer,2014Adhikari,2015More} as an alternative to radii based on overdensity to minimize the impact of such issues. The splashback radius is defined as the radius where particles reach the apocenter of their first orbit, observationally characterized by the logarithmic slope of the density distribution showing a deep drop \citep{2015More}. In the case of spherical symmetry, this radius indicates a distinct boundary between infalling matter and matter that is already orbiting within the halo. Given a measured splashback radius, one can then measure observables within this radius that are expected to correlate with mass, such as weak lensing shear, without the added degeneracy associated with overdensity masses \citep{2024Giocoli}. Additionally, the splashback radius alone can be used as a mass proxy if one assumes a functional form for the cluster mass profile \citep{2023Contigiani}. The splashback radius could also provide a competitive constraint on cosmological parameters by comparing with results from simulation because it breaks the degeneracy between $\Omega_{\rm M}$ and $\sigma_8$ \citep{2024Haggar,2024Mpetha,2025Mpetha}. 

Many studies using cosmological simulations have investigated the physical interpretation of the splashback radius. \cite{2021ONeil} confirmed the detection of the splashback feature in dark matter, gas, and satellite galaxies in hydrodynamical simulations such as IllustrisTNG. Using the Hydrangea simulations, \cite{2021Contigiani} found correlations between the splashback radius, the direction of the filaments, and the orientation of brightest cluster galaxies. \cite{2023Shin} demonstrated that the splashback radius is a clean tracer of halo growth over its past dynamical time. More recently, \cite{2025Walker} showed that the splashback feature traced by stellar mass closely coincides with that of the dark matter distribution in GIZMO simulations from the Three Hundred Project. In addition, several models have been proposed to predict the splashback radius \cite{2014Diemer,2014Adhikari,2015More,2016Shi,2017Mansfield,2017Diemer,2023Diemer}. Across these studies, the mass accretion rate of dark matter halos consistently exhibits a tight correlation with the splashback radius. This establishes the splashback radius as a powerful probe of cluster assembly history.

The first significant observational detection of splashback radius comes from \cite{2016More} using cross-correlation between the redMaPPer Sloan Digital Sky Survey (SDSS) clusters \citep{2014Rykoff} and SDSS galaxies. Following this work, detections of splashback radii have been made by multiple investigators using different galaxy cluster surveys. On the optical/infrared side, detections of splashback radii were shown in redMaPPer Dark Energy Survey (DES) clusters \citep{2017Baxter,2018Chang}, Hyper Suprime-Cam (HSC) CAMIRA clusters \citep{2018Nishizawa,2020Murata}, Adaptive Matched Identifier of Clustered Objects in Kilo-Degree Survey Data Release 3 (AMICO KiDS-DR3) clusters \citep{2024Giocoli}, and Dark Energy Spectroscopic Instrument (DESI) Legacy Imaging Surveys (LS) clusters \citep{2024Xu} using cluster weak lensing and/or cluster-galaxy cross-correlation. For surveys based on Sunyaev–Zeldovich (SZ) detections, the splashback radius has been detected in the Planck cluster catalog \citep{2019Zurcher} and a catalog combining the Atacama Cosmology Telescope (ACT) and the South Pole Telescope (SPT) cluster catalogs \citep{2019Shin,2021Shin}, all using cluster-galaxy cross-correlation and weak lensing. For X-ray based surveys, the splashback radius has been observed in the weak lensing of the Cluster Lensing And Supernova survey with Hubble (CLASH) clusters \citep{2017Umetsu} and the Canadian Cluster Comparison Project (CCCP) clusters \citep{2019Contigiani}. \cite{2021Bianconi} presents the splashback feature in a stacked luminosity density profile. The eROSITA Final Equatorial-Depth Survey (eFEDS) also shows the splashback radius based on cross-correlation measurement in \cite{2023Rana}. 
 
All of these previous works utilize cluster catalogs that are limited to $z < 1$. The redshift, mass range \citep[$0.1<z<2$, $M> 1\times10^{13} M_\odot$,][]{2024Thongkham,2024Thongkhamb}, and large sample size of MaDCoWS2 provide the first opportunity to measure the splashback feature for a uniform cluster sample out to an epoch where changes in cluster scales are expected to be significant.

In this work, we use cross-correlation between MaDCoWS2 cluster candidates and WISE galaxies to investigate the MaDCoWS2 cluster profiles and detect the splashback radius at $0.4<z<1.65$. We also explore the dependence of the splashback radius on redshift and S/N$_{\rm P}$ (Signal-to-noise under the assumption that the noise can be approximated as Poissonian) for the MaDCoWS2 sample. This is the first step towards an independent assessment of the radial boundary and characteristic mass limit of the MaDCoWS2 clusters as a function of their redshift and S/N$_{\rm P}$. 

Section \ref{sec:data} describes the cluster and galaxy catalogs used in this paper. Section \ref{sec:measurement} details the methodology, explaining the surface density measurement, the model used, and the model fitting method. We present the density profiles based on the cross-correlation and the splashback radius measurements in Section \ref{sec:result and discussion}, while also discussing the implications of these results. We summarize our conclusions in Section \ref{sec:conclusion}. 

Throughout this paper, we assume a flat $\Lambda$CDM cosmology from \cite{2020Planck} with $\Omega_m =0.315$ and $H_0 = 67.4$ km s$^{-1}$ Mpc$^{-1}$. In Tables and Figures however, we report quantities with a dimensionless Hubble parameter $h$ where $H_0=100\;h$\;km$^{-1}$s$^{-1}$Mpc$^{-1}$ to facilitate comparison with other studies. Distances in the comoving and proper frames are denoted by cMpc and pMpc respectively. $R$ and $r$ denote the projected and three-dimensional radii, respectively.

\section{Data} \label{sec:data}
\subsection{MaDCoWS2 clusters}
The MaDCoWS2 catalog \citep{2024Thongkham,2024Thongkhamb} consists of $133,036$ cluster candidates with S/N$_{\rm P}$ $\geq 5$ at $0.1<$ z $<2$. The galaxy clusters in the catalog are detected using CatWISE2020-selected galaxies \citep{2020Eisenhardt,2021Marocco}. Photometric redshift probability distribution functions (PDFs) are derived for each galaxy using $g$, $r$, $z$, W1, and W2 data from the the ninth data release of Dark Energy Camera Legacy Survey \citep[DECaLS DR9, see][]{2019Dey} and from CatWISE2020 (see \S \ref{sec:galaxy}). The search identifies the overdensities of galaxies in three dimensions (sky positions and photometric redshifts) using the PZWav algorithm \citep{2014Gonzalez,2019Euclid,2023Werner,2024Thongkham,2024Thongkhamb,2025Bhargava}. This catalog covers $6498$ ($8436$ without masking) deg$^2$ spanning the majority of the DECaLS area. 

\begin{figure}[htbp]
\centering
\includegraphics[width=0.9\columnwidth]{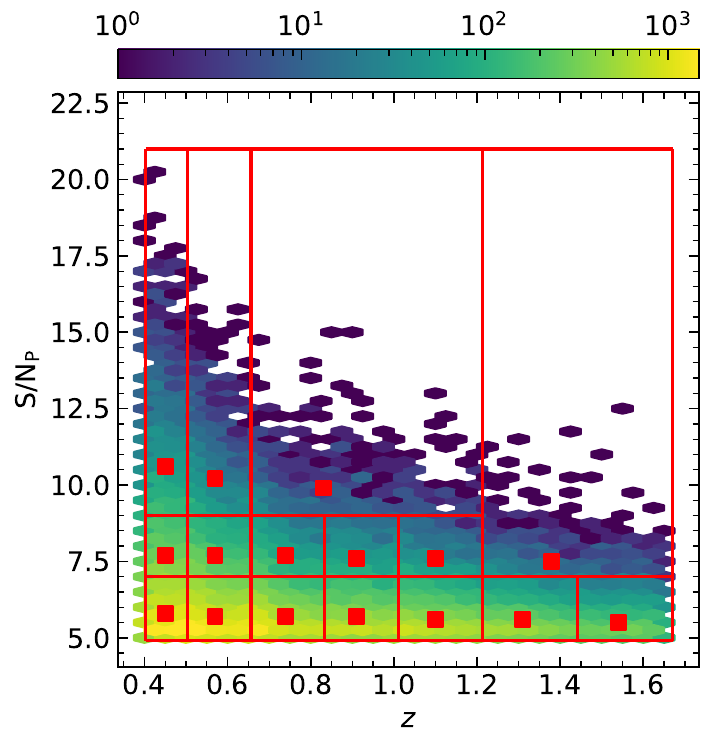}
\caption{Sample selection of the cluster candidates used in the analysis of this paper. The cluster sample is displayed in hexagonal bins with a bin size of $0.05$ in cluster redshift on the x-axis and $0.5$ in S/N$_{\rm P}$ on the y-axis. S/N$_{\rm P}$ and $z$ are the signal-to-noise ratio and photometric redshift of the clusters. Bin colors indicate the number of clusters. The solid red lines display the binning scheme for subsamples shown in Table \ref{Tab:cluster sample}. The red squares indicate $\overline{\text{S/N}}_{\rm P}$ and $\overline{z}$ of the subsamples.}
\label{fig:cluster sample}
\end{figure}

Our analysis considers cluster candidates divided into subsamples according to their properties, specifically photometric redshift and S/N$_{\rm P}$, where S/N$_{\rm P}$ is defined as the signal-to-noise ratio of high-density peaks in a three-dimensional galaxy density cube relative to the noise estimated from a random density cube assuming that it follows a Poisson distribution. We divide the cluster sample into multiple subsamples with photometric redshift and S/N$_{\rm P}$ bins shown in Figure \ref{fig:cluster sample}, within which we calculate the mean cluster-galaxy cross-correlations to estimate the projected galaxy density around each cluster subsample. The number of cluster candidates in each subsample selection is provided in Table \ref{Tab:cluster sample}. This subdivision is designed to facilitate comparison with \citet{2024Trudeau}.

We focus on cluster candidates at $0.4<$ z $<1.65$ in this work. This redshift range ensures a low level of systematic and statistical uncertainties in the density profile measurements. Outside this redshift range, the identification of a splashback radius with well-constrained uncertainty from the galaxy density profile becomes harder. Our cut based on absolute W1 magnitude results in a smaller number of galaxies at z $<0.4$. At $z>1.65$, the number of clusters is low due to the photometric limits of the galaxy and cluster catalogs. This leads to a higher variance in the galaxy density outside of the core profile and prevents a robust measurement of the splashback radius in these redshift ranges.

 \begin{deluxetable}{cccc}
 \tablecaption{Cluster counts in each bin, organized by redshift (first column) and S/N$_{\rm P}$ (three other columns.) \label{Tab:cluster sample}}
 \tablehead{\colhead{$z$} & \colhead{$5\leq$ S/N$_{\rm P}$ $<7$} & \colhead{$7\leq$ S/N$_{\rm P}$ $<9$} & \colhead{S/N$_{\rm P}$ $\geq 9$}}
 \startdata
 0.4-0.5 & 15613 & 4724 & 2065 \\
 0.5-0.65 & 19382 & 4139 & 1088 \\
 0.65-0.83 & 15667 & 2255 & \nodata \\
 0.83-1.0 & 11439 & 1381 & \nodata \\
 1.0-1.2 & 10862 & 917 & \nodata \\
 1.2-1.43 & 8108 & \nodata & \nodata \\
 1.43-1.65 & 4843 & \nodata & \nodata \\
 \hline
 1.2-1.65 & \nodata & 637 & \nodata \\
 0.65-1.2 & \nodata & \nodata & 612 \\
 \enddata
 \end{deluxetable}

\subsection{Galaxy catalogs} \label{sec:galaxy}
We use the galaxy catalogs built for the MaDCoWS2 cluster search as the primary input of the galaxy cross-correlation analysis. We refer readers to \cite{2024Thongkham} for a detailed description of the process, which we summarize below.

The positions of the galaxies in this analysis are from the infrared CatWISE2020 catalog. We employ the same filtering and masking procedures as in MaDCoWS2 on the galaxy sample. This includes filtering artifacts based on artifact flags from CatWISE2020 and DECaLS, removing stars based on $r-z$ and $z-W1$ colors, and removing galaxies using large galaxy masks and bright star masks. We also reject galaxies with a low probability of being a match between CatWISE2020 and DECaLS. The galaxy catalog used in MaDCoWS2 also requires DECaLS detections, which hinders the detection S/N for the highest-z clusters. To prevent faint galaxies from significantly contributing to the correlation estimate of the subsample at low redshift, we employ an absolute magnitude cut of $M=-23.01$ AB in the WISE W1 band, equivalent to an apparent magnitude of $W1=18.74$ AB at $z=0.4$ and $W1=23$ AB at $z=2$. 

The galaxy selection can impact the redshift dependence of the derived radii if the splashback radius is dependent upon either galaxy mass or galaxy color. Selections based on artifact flags, stellar rejection, and masking should not introduce redshift or mass-dependent biases to our analysis. Selecting galaxies down to a constant absolute magnitude should minimize any dependence on galaxy mass since it ensures similar galaxies are being looked at in each redshift bin. Requiring 5-band photometry will have some impact at high redshift. Due the magnitude limit of our galaxy catalog, we would start losing a lot of galaxies only at the highest redshift bins. Nevertheless, it has also been argued in \cite{2020Murata} that the splashback radius constraint is not strongly dependent on color up to  $z<1$, so we do not expect the bias from our selection to be strong.

\section{Galaxy number density} \label{sec:measurement}
\subsection{Measurement} \label{subsec:measurement}
We measure the galaxy surface density around our clusters using cluster galaxy cross-correlation similar to the correlation performed in \cite{2019Shin,2021Shin} and \cite{2020Murata}. We use a modified version of the Landy-Szalay estimator \citep{1993Landy} to calculate the two-point cross correlation between MaDCoWS2 clusters and the galaxy sample in \S \ref{sec:galaxy}. The estimator follows the equation
\begin{equation}
\label{eq:omegaR_cal}
    \omega (\theta) = \frac{D_cD_g-D_cR_g-R_cD_g+R_cR_g}{R_cR_g},
\end{equation}
\noindent where we compute normalized pairs of objects as a function of angular separation. The normalization is calculated with respect to the total number of pairs processed. $D$ and $R$ denote objects in the true and random catalogs, respectively, while the subscripts $c$ and $g$ refer to clusters and galaxies, respectively. We use a random cluster catalog with $100$ times the number of galaxy clusters in our sample. The random galaxy data has $4$ times the surface number density of the galaxy catalog. Survey masks are included when generating the random catalogs.

Each cluster subsample is divided into another set of subsamples with redshift binning of $\Delta z=0.025$. For each of these subsample bins, we use \texttt{treecorr} \citep{2004Jarvis,2015Jarvis} to estimate angular two-point cross-correlations that we convert into comoving radial correlations using the angular diameter distance at the midpoint of the redshift bin. The correlation is computed over $40$ comoving radial bins evenly divided in logarithmic space from $0.2$ to $5$ $h^{-1}$ Mpc for all subsamples except for those with S/N$_{\rm P}\geq9$. By visual inspection and initial model fitting, we find that the infalling profile of those subsamples is not fully captured by the default radius limits of $5$ $h^{-1}$ Mpc. Thus, we use $50$ bins from $0.2$ to $8$ $h^{-1}$ Mpc instead for the S/N$_{\rm P}\geq9$ subsamples to ensure their entire halo profiles are captured by our analysis. 

Across all subsamples, the maximum radial bin sizes are smaller than the median projected separation between clusters, $S_{\rm proj}$. $S_{\rm proj}$ varies from $5.2$ Mpc at $5\leq$ S/N$_{\rm P}$$< 7$ and $0.4<z<0.5$ to $39.5$ Mpc at S/N$_{\rm P}\geq 9$ and $0.65<z<1.2$. The fraction of clusters overlapping in projection, defined as the fraction of overlap within 2 $h^{-1}$ Mpc, correspondingly decreases from 11$\%$ to 1$\%$ across these bins. Consequently, we do not expect projected cluster overlap to significantly affect our splashback radius measurements.

From the projected comoving radius $R$, we calculate the mean 2-point cross-correlation $\omega (R)$ for a larger redshift bin with $\Delta z=0.3$ by averaging the correlations using weights from the number of clusters in each redshift bin following 
\begin{equation}
\label{eq:omegaR}
    \omega (R) = \frac{\sum_i N_{cl,i} \omega (R,z_i)}{\sum_i N_{cl,i}},
\end{equation}
where $N_{cl,i}$ is the number of cluster in each $\Delta z=0.025$ redshift bin and $\omega (R,z_i)$ is the correlation of that small redshift bin. The mean subtracted projected galaxy profile $\Sigma_g(R)$ then can be obtained by 

\begin{equation}
\label{eq:sigmagR}
    \Sigma_g(R) = \overline{\Sigma}_g\omega (R),
\end{equation}
where $\overline{\Sigma}_g$ is the average surface number density of the sample of galaxy described by

\begin{equation}
\label{eq:sigmag}
    \overline{\Sigma}_g = \frac{\sum_i N_{cl,i} \overline{\Sigma}_{g,i}}{\sum_i N_{cl,i}}
\end{equation}

where $\overline{\Sigma}_{g,i}$ is the galaxy average surface number density in each small redshift bin calculated by dividing the number of galaxies within the bin by the surface area in $h^{-2}$ cMpc$^2$. We estimate the covariance matrix of the cross-correlation using a jackknife resampling method \citep{2009Norberg}, available as a part of \texttt{treecorr}, with $200$ patches of approximately equal size with each patch representing a subdivision of the survey area. The jackknife resampling estimates the covariance matrix from correlations measured when one patch at a time is excluded from the sample with this process repeating until the total number of patches has been reached. The matrix can be described by 

\begin{equation}
\label{eq:covariance matrix}
    C = \frac{N_{\text{patch}}-1}{N_{\text{patch}}} \sum_i (\omega_i-\overline{\omega})^T(\omega_i-\overline{\omega}).
\end{equation}

\subsection{Model} \label{sec:Model}
 We model the galaxy surface density based on a line-of-sight integration of a spherical density profile proposed in \citet[][hereafter D23]{2023Diemer}. Previous works traditionally relied on a similar profile from \citet[][DK14]{2014Diemer} to fit their density profiles. Both DK14 and D23 can be written as an addition of a profile composed of orbiting galaxies ($\rho_{\rm orbit}$) and another profile composed of infalling galaxies ($\rho_{\rm infall}$). These profiles are designed to encapsulate the features of dark matter halos out to large radii where material falling into the gravitational well of the clusters can be described through an infalling power-law term. D23 improves on the model in DK14, which is traditionally used as a fitting function for splashback analysis in many previous works, by providing more accurate fitting using more free parameters. In this work, we use a modified version of D23 which can be described as

\begin{equation}
\label{eq:rho}
    \rho(r) = \rho_{\text{orbit}}(r) + \rho_{\text{infall}}(r)
\end{equation}
where
\begin{align}
\label{eq:rho orbit}
    \rho_{\text{orbit}}(r) & = \rho_s \text{exp} \left( -\frac{2}{\alpha}\left[\left(\frac{r}{r_s}\right)^{\alpha}-1\right] \right. \\ \nonumber & \left. - \frac{1}{\beta}\left[\left(\frac{r}{r_t}\right)^{\beta} - \left(\frac{r_s}{r_t}\right)^{\beta}\right] \right),   
\end{align}
\begin{equation}
\label{eq:rho infall}
    \rho_{\text{infall}}(r) = \delta_1 \left[\left(\frac{\delta_1}{\delta_{\text{max}}}\right)^{2} + \left(\frac{r}{r_{\text{pivot}}} \right)^{2s} \right]^{-1/2}
\end{equation}

and the other terms are defined in Table \ref{Tab:prior}. The modification removes the mean density of the Universe ($\rho_m$) from the infalling term of D23, as our analysis involves the galaxy density profile instead of the matter density profile measured by weak lensing.

Similar to \cite{2019Zurcher} and \cite{2020Murata}, we obtain a surface density profile using 

\begin{equation}
\label{eq:sigmagRint}
    \Sigma_g(R) = \frac{1}{l_{\text{max}}}\int_0^{l_{\text{max}}}dl \;\rho  \left(\sqrt{R^2+l^2}\right),
\end{equation}
where $R$ is the projected radius and $l$ is the projected distance to the cluster. We set $l_{\text{max}}=40 h^{-1}$ Mpc and the pivot radius $r_{\text{pivot}}=1.5 h^{-1}$ Mpc, leaving $8$ free parameters ($\alpha, \beta, \rho_s, r_s, r_t, \delta_1, \delta_{\text{max}},$ and $s$) in our fitting process.

\subsection{Model fitting} \label{sec:Model fitting}
Following the standard methodology of model fitting for a splashback analysis \citep{2016More,2019Shin,2021Shin,2020Murata}, we fit the model described in \S\ref{sec:Model} using a Bayesian approach. The likelihood function is described as

\begin{equation}
\label{eq:likelihood}
    \text{ln}\;L\left[\vec{d}\mid \vec{m}({\theta})\right] = -\frac{1}{2}\left[\vec{d}-\vec{m}(\vec{\theta})\right]^\text{T} C^{-1} \left[\vec{d}-\vec{m}(\vec{\theta})\right],
\end{equation}
where $\vec{d}$ is the data vector, $\vec{m}$ is the model, $\vec{\theta}$ is the model parameter, and $C$ is the covariance matrix. The data vector is the projected galaxy density profile measured in $40$ or $50$ radial bins for subsamples below or above S/N$_{\rm P}=9$, respectively. We then use a posterior

\begin{equation}
\label{eq:postirioer}
    \text{ln}\;P\left[\vec{\theta}\mid \vec{d}\right] = \text{ln}\;L\left[\vec{d}\mid\vec{m}(\vec{\theta})\text{Pr}(\vec{\theta})\right],
\end{equation}
where $\text{Pr}(\vec{\theta})$ is the prior of the model parameters. 

We provide the priors used in our fitting process in Table \ref{Tab:prior}. We set generous ranges for our prior to ensure that the best fit can be achieved through our fitting process, considering the difference in the mass and redshift range of our sample compared with other cluster and galaxy surveys that have been previously investigated. We also require that $r_t$ is larger than $r_s$ following D23.

\begin{deluxetable*}{ccc}
\tablecaption{Prior range for each model parameter in \S \ref{sec:Model} \label{Tab:prior}}
\tablehead{\colhead{Parameters} & \colhead{Prior} & \colhead{Description}}
\startdata
$\log \alpha$ & $[-3,1]$ & Steepening parameter \\
$\log \beta$ & $[-3,1]$ & Truncation sharpness \\
$\log \left( \frac{\rho_s}{[h^3 \rm{Mpc^{-3}}]} \right)$ & $[-2,5]$ & Overdensity at scale radius\\
$\log \left( \frac{r_s}{[h^{-1}\rm{Mpc}]} \right)$ & $\log [0.2,5(8)]$ & Scale radius \\
$\log \left( \frac{r_t}{[h^{-1}\rm{Mpc}]} \right)$ & $\log [0.2,5(8)]$ & Truncation radius  \\
$\log \left( \frac{\delta_1}{[h^3 \rm{Mpc^{-3}}]} \right)$ & $[-2,4]$ & Infalling normalization   \\
$\log \left( \frac{\delta_{\text{max}}}{[h^3 \rm{Mpc^{-3}}]} \right)$ & $[-1,4]$ & Infalling central overdensity \\
$s$ & $[0.1,10]$ & Infalling slope\\
\enddata 
\tablecomments{Radial ranges in parenthesis indicate a different upper limit for subsamples with S/N$_P\geq9$.}
\end{deluxetable*}

We employ an Affine Invariant Markov Chain Monte Carlo (MCMC) ensemble sampler \citep{2010Goodman} from \texttt{emcee} \citep{2013Foreman} to sample the posterior of our parameters, as described in Equation \ref{eq:postirioer}. We use 50 chains and 6000 steps with a burn-in period of 2500 steps. One set of parameters is retained every 50 steps to construct the final sample, from which we derive the statistical results reported in \S\ref{sec:result and discussion}. We ensure convergence by inspecting the time streams of each parameter. 

\section{Results and Discussion} \label{sec:result and discussion}

We cross-correlate the galaxy sample in \S \ref{sec:galaxy} with MaDCoWS2 clusters, divided into subsamples by redshift and S/N$_{\rm P}$. The cross-correlations are transformed into galaxy surface densities and fitted to a density profile using the methodology described in \S \ref{sec:measurement}. In this section, we investigate the halos of MaDCoWS2 cluster candidates, obtain their splashback radii, and compare our results with those in the literature.

\subsection{Density Profiles} \label{sec:dens prof}

\begin{figure*}[htbp]
\centering
\includegraphics[width=0.88\textwidth]{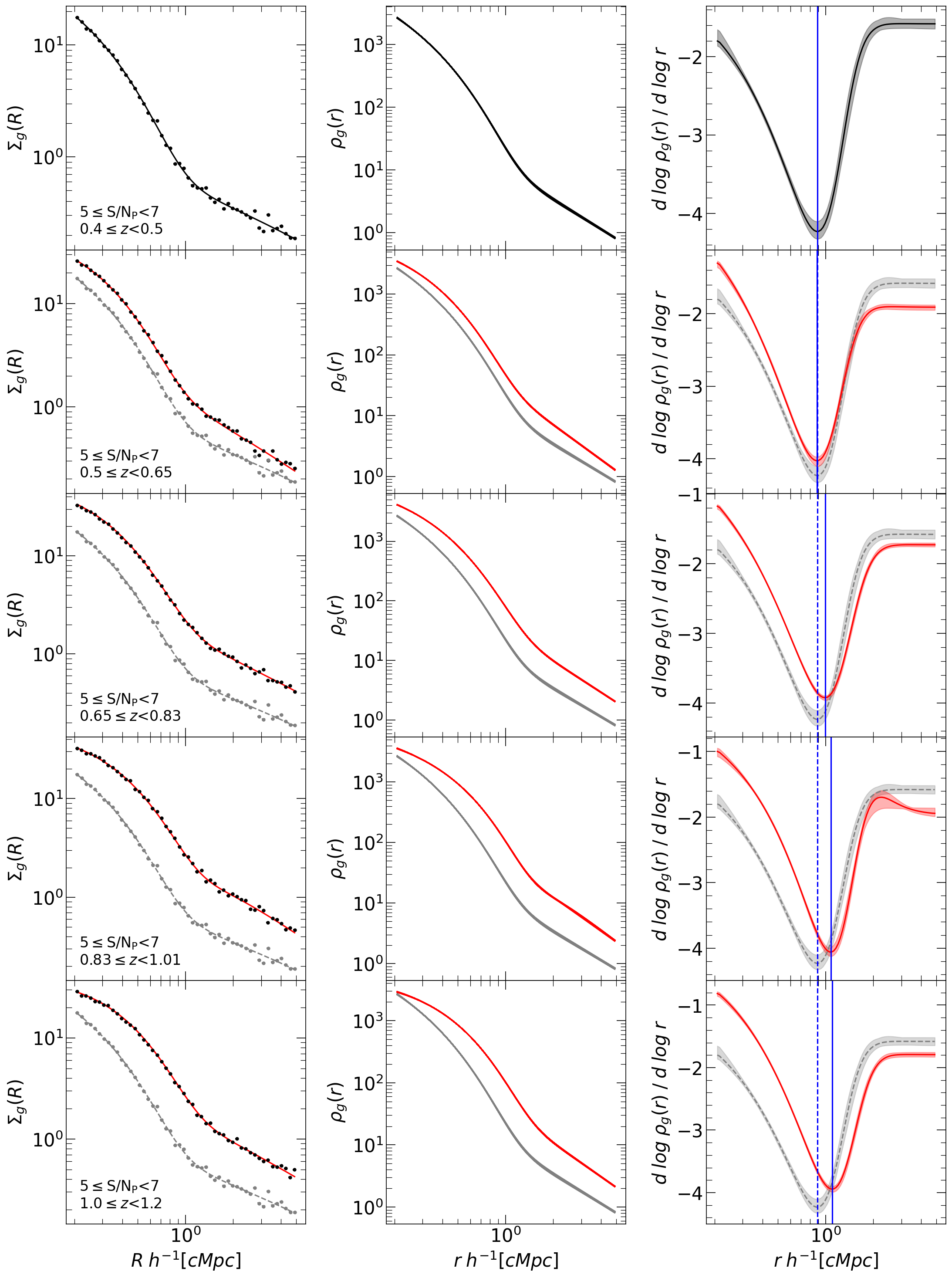} 
\caption{\label{fig:dens1} Galaxy surface density versus comoving projected radius (left column), galaxy density versus comoving three-dimensional radius (middle column), and logarithmic derivative of galaxy density versus comoving three-dimensional radius (right column) of MaDCoWS2 clusters with $5\leq$ S/N$_{\rm P}$ $<7$. Each row displays a different redshift subsample as listed in Table \ref{Tab:cluster sample}. The surface density measurements as described in \S \ref{subsec:measurement} are shown as data points in the left column. The solid lines in the left column are the best fit to galaxy surface density from the model fitting scheme in \S \ref{sec:Model fitting}. The solid lines in the right column show the three-dimensional splashback radii ($r_{\rm sp}$). For reference, dashed lines show the profiles and splashback radius of the lowest redshift bin in the higher redshift subsamples. We input all fit parameters in our MCMC samples to the model in \S \ref{sec:Model} and generate the curves in the middle columns. The curves in the right column correspond to the best-fit parameters, and to the 0th and 100th percentiles of the parameter distribution.}
\end{figure*}

\begin{figure*}[htbp]
\addtocounter{figure}{-1}
\centering
\includegraphics[width=0.88\textwidth]{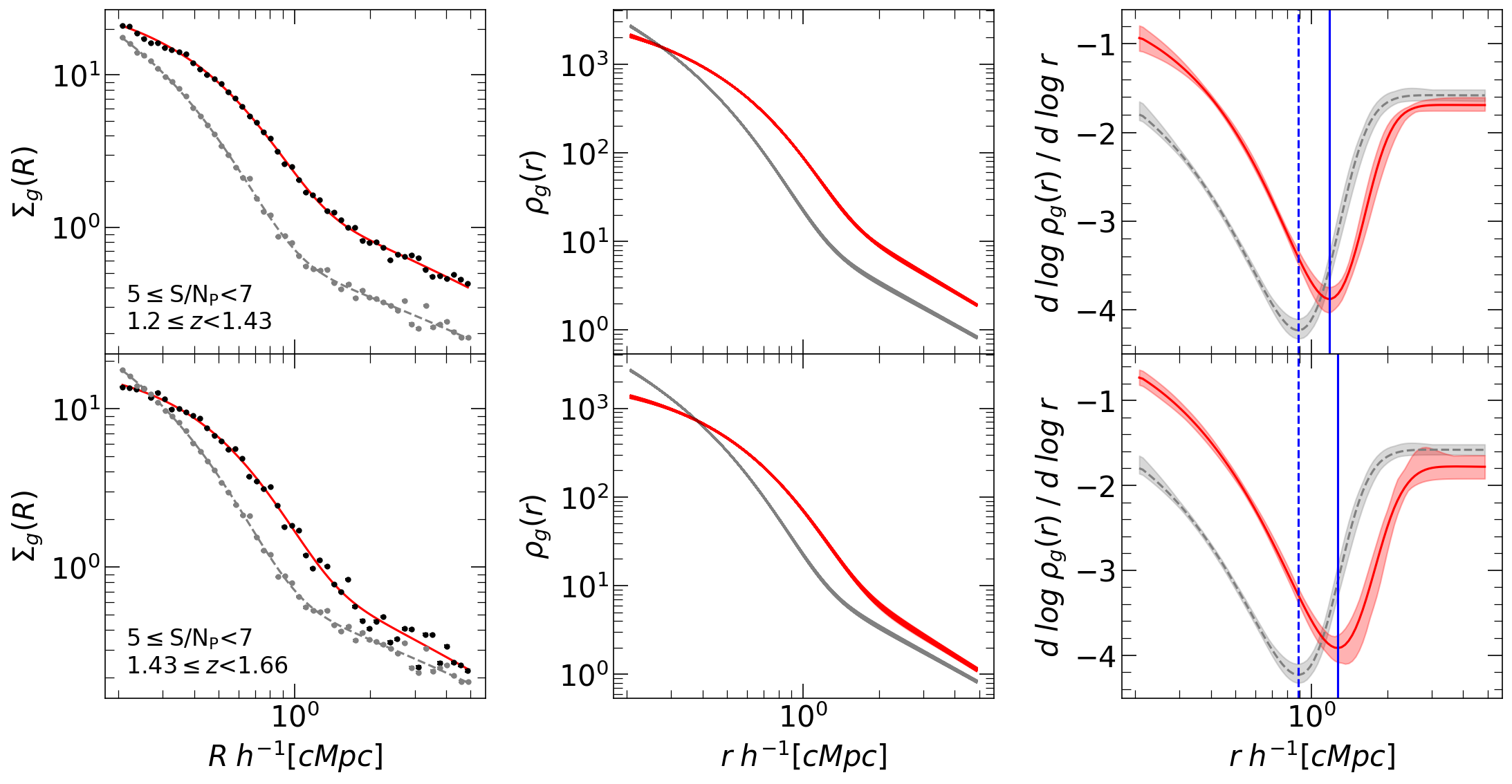} 
\caption{Continued)}
\end{figure*}

\begin{figure*}[htbp]
\centering
\includegraphics[width=0.88\textwidth]{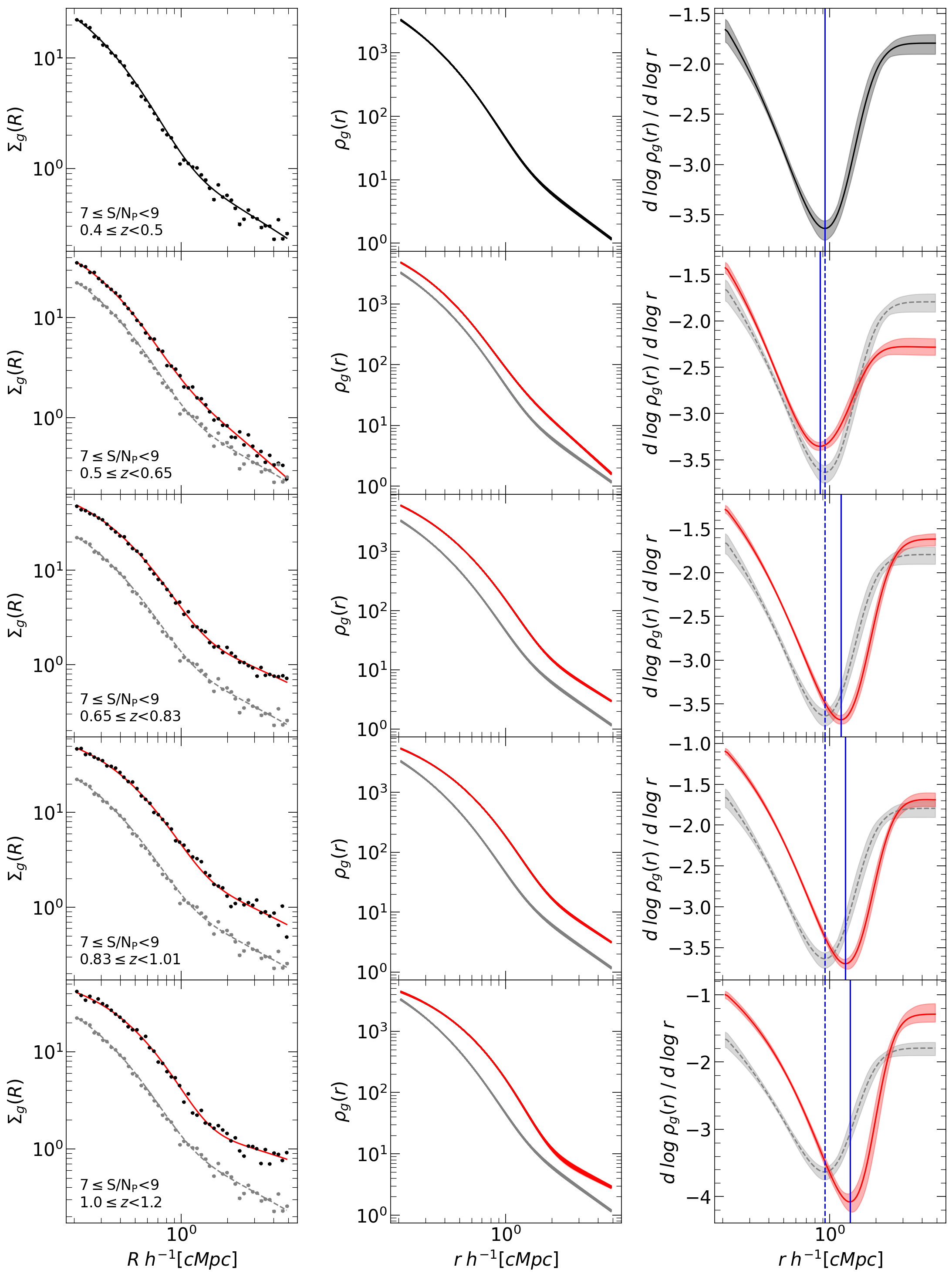} 
\caption{\label{fig:dens2} Galaxy surface density (left column), galaxy density (middle column), and logarithmic derivatives of galaxy density (right column) similar to Figure \ref{fig:dens1}, but for subsamples with $7\leq$ S/N$_{\rm P}$ $<9$.}
\end{figure*}

\begin{figure*}[htbp]
\centering
\includegraphics[width=0.88\textwidth]{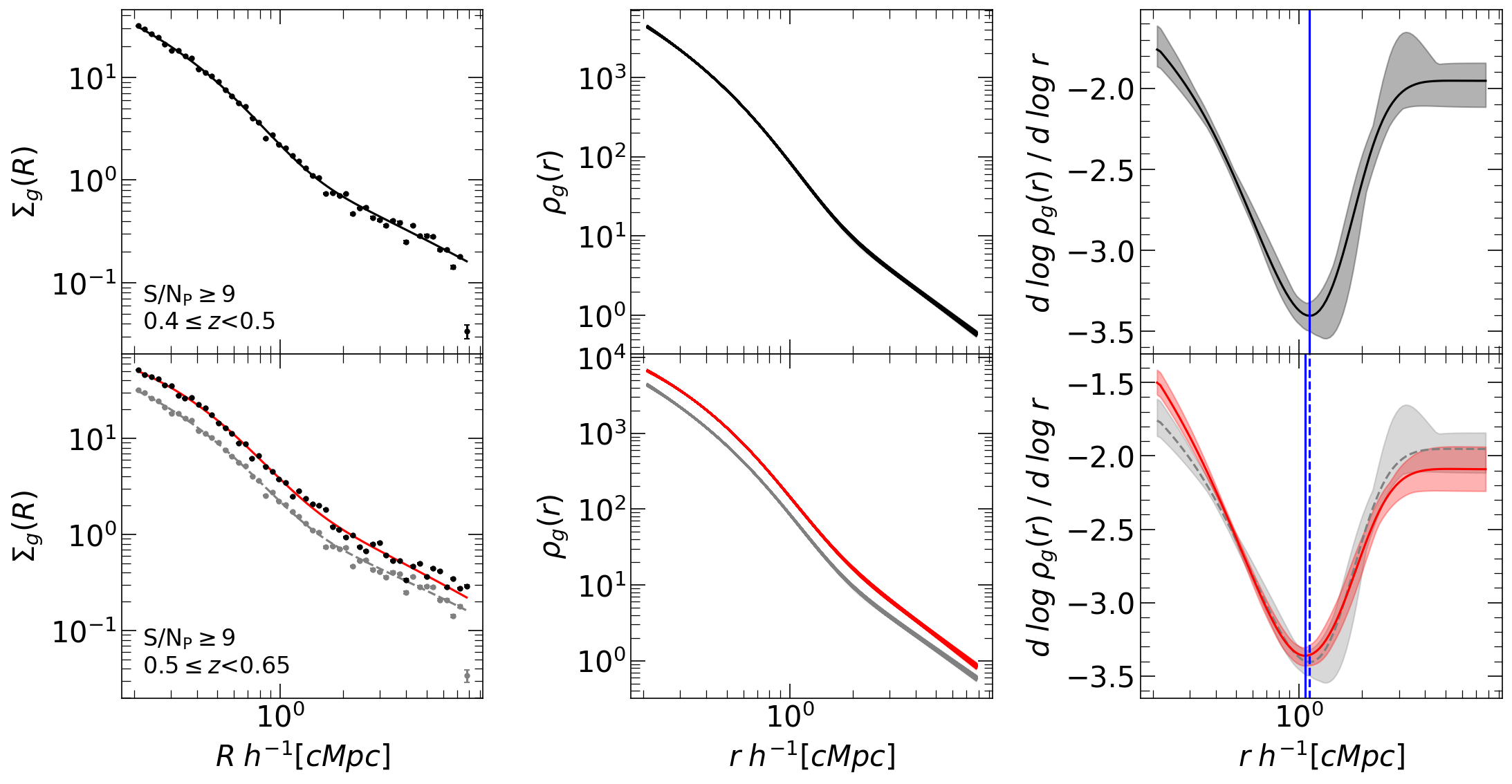} 
\caption{\label{fig:dens3} Galaxy surface density (left column), galaxy density (middle column), and logarithmic derivatives of galaxy density (right column) similar to Figure \ref{fig:dens1}, but for subsamples with S/N$_{\rm P}$ $\geq 9$.}
\end{figure*}

\begin{figure*}[htbp]
\centering
\includegraphics[width=0.88\textwidth]{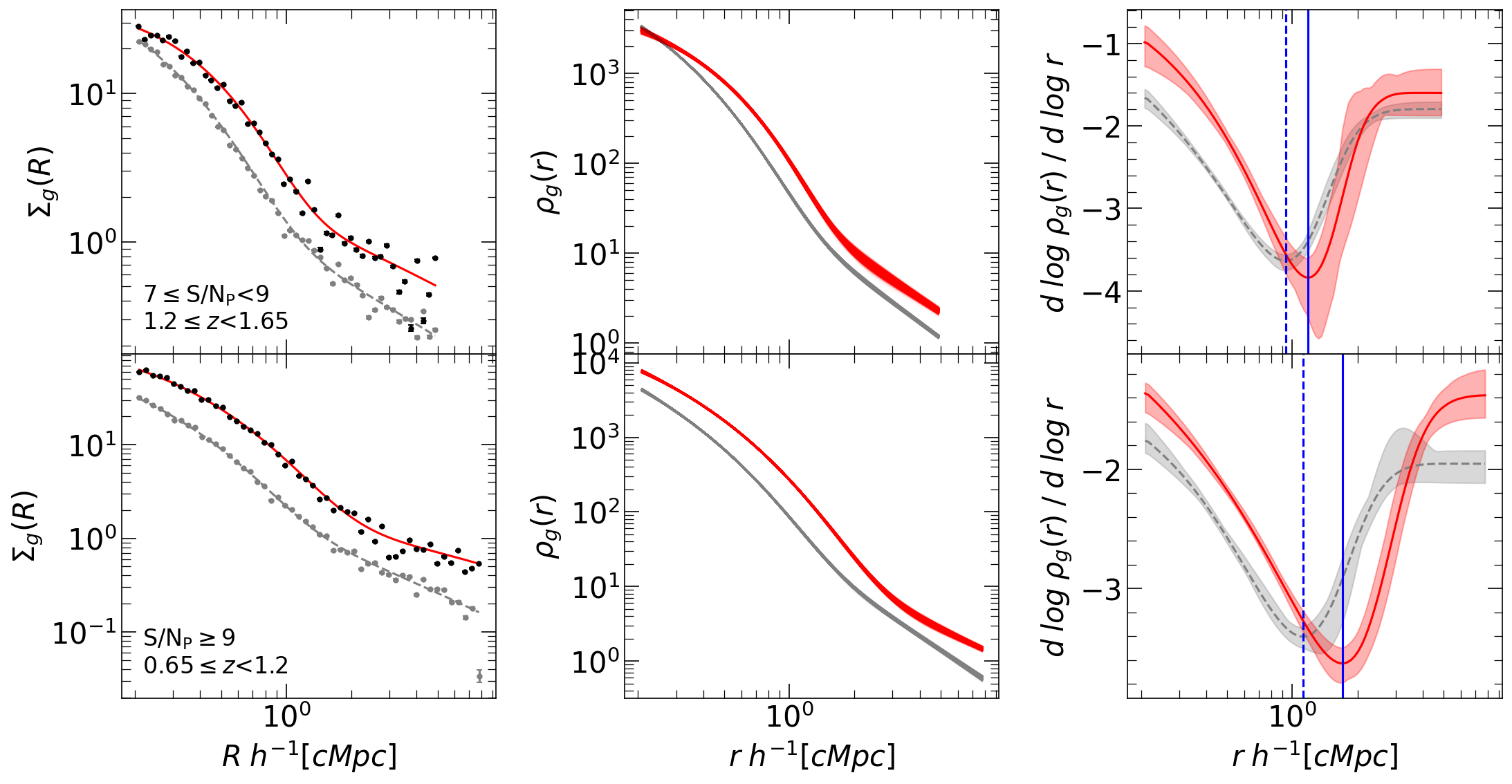} 
\caption{\label{fig:dens4} Galaxy surface density (left column), galaxy density (middle column), and logarithmic derivatives of galaxy density (right column) similar to Figure \ref{fig:dens1}, but for $7\leq$ S/N$_{\rm P}$ $<9$ and S/N$_{\rm P}$ $\geq9$ with larger redshift binning as described in Table \ref{Tab:cluster sample}.}
\end{figure*}

\begin{deluxetable*}{llrrrrrrrr}
\tablecaption{Best-fit Parameters for Cluster Subsamples\label{tab:fit_results}, Measured in Comoving Units}
\tablewidth{0pt}
\tablehead{
\colhead{S/N$_{\rm P}$} & 
\colhead{$z$} & 
\colhead{$\log \alpha$} & 
\colhead{$\log \beta$} & 
\colhead{$\log \rho_{s}$} & 
\colhead{$\log r_s$} & 
\colhead{$\log r_t$} & 
\colhead{$\delta_1$} & 
\colhead{$\delta_{\rm max}$} & 
\colhead{s}
}
\startdata
$5 \leq$ S/N$_{\rm P}$$< 7$ & $0.4 - 0.5$   & $0.05^{+0.02}_{-0.04}$ & $-1.39^{+0.62}_{-0.78}$ & $2.68^{+0.08}_{-0.02}$ & $-0.37^{+0.01}_{-0.03}$ & $0.17^{+0.37}_{-0.33}$ & $0.73^{+0.01}_{-0.01}$ & $2.43^{+1.06}_{-0.91}$ & $1.58^{+0.02}_{-0.02}$ \\
                           & $0.5 - 0.65$  & $0.01^{+0.03}_{-0.03}$ & $-0.02^{+0.17}_{-0.14}$ & $3.03^{+0.08}_{-0.10}$ & $-0.41^{+0.04}_{-0.04}$ & $-0.08^{+0.25}_{-0.13}$ & $1.09^{+0.01}_{-0.01}$ & $1.78^{+0.10}_{-0.08}$ & $1.91^{+0.01}_{-0.01}$ \\
                           & $0.65 - 0.83$ & $0.02^{+0.02}_{-0.02}$ & $-0.02^{+0.07}_{-0.08}$ & $2.97^{+0.08}_{-0.10}$ & $-0.34^{+0.04}_{-0.03}$ & $-0.06^{+0.14}_{-0.12}$ & $1.20^{+0.01}_{-0.01}$ & $2.50^{+0.12}_{-0.64}$ & $1.73^{+0.01}_{-0.01}$ \\
                           & $0.83 - 1.0$  & $0.12^{+0.02}_{-0.02}$ & $-0.35^{+0.08}_{-0.10}$ & $2.75^{+0.07}_{-0.05}$ & $-0.22^{+0.02}_{-0.03}$ & $-0.04^{+0.26}_{-0.11}$ & $1.39^{+0.01}_{-0.01}$ & $1.35^{+0.04}_{-0.03}$ & $1.97^{+0.02}_{-0.02}$ \\
                           & $1.0 - 1.2$   & $0.06^{+0.01}_{-0.01}$ & $0.05^{+0.05}_{-0.05}$ & $2.79^{+0.08}_{-0.09}$ & $-0.27^{+0.03}_{-0.03}$ & $0^{+0.15}_{-0.11}$ & $1.25^{+0.01}_{-0.01}$ & $2.41^{+0.07}_{-0.65}$ & $1.79^{+0.01}_{-0.01}$ \\
                           & $1.2 - 1.43$  & $0.15^{+0.04}_{-0.20}$ & $-0.51^{+0.74}_{-0.26}$ & $2.48^{+0.10}_{-0.05}$ & $-0.19^{+0.02}_{-0.04}$ & $0.07^{+0.40}_{-0.18}$ & $1.15^{+0.01}_{-0.01}$ & $1.97^{+0.13}_{-0.13}$ & $1.69^{+0.02}_{-0.02}$ \\
                           & $1.43 - 1.65$ & $0.05^{+0.02}_{-0.02}$ & $0.22^{+0.51}_{-0.25}$ & $2.57^{+0.04}_{-0.09}$ & $-0.28^{+0.05}_{-0.01}$ & $0.43^{+0.17}_{-0.20}$ & $0.97^{+0.02}_{-0.02}$ & $2.02^{+0.19}_{-0.73}$ & $1.78^{+0.04}_{-0.04}$ \\
$7 \leq$ S/N$_{\rm P}$$< 9$ & $0.4 - 0.5$   & $-0.13^{+0.07}_{-0.05}$ & $-0.23^{+0.23}_{-0.35}$ & $2.89^{+0.24}_{-0.15}$ & $-0.40^{+0.05}_{-0.10}$ & $0.08^{+0.39}_{-0.28}$ & $0.99^{+0.01}_{-0.01}$ & $2.65^{+1.00}_{-0.65}$ & $1.79^{+0.03}_{-0.02}$ \\
                           & $0.5 - 0.65$  & $-0.07^{+0.02}_{-0.02}$ & $-0.08^{+0.08}_{-0.08}$ & $3.20^{+0.11}_{-0.22}$ & $-0.42^{+0.09}_{-0.05}$ & $0.15^{+0.38}_{-0.35}$ & $1.38^{+0.01}_{-0.01}$ & $1.80^{+0.05}_{-0.05}$ & $2.29^{+0.02}_{-0.02}$ \\
                           & $0.65 - 0.83$ & $-0.12^{+0.02}_{-0.02}$ & $-0.13^{+0.04}_{-0.05}$ & $2.96^{+0.11}_{-0.12}$ & $-0.26^{+0.04}_{-0.04}$ & $-0.03^{+0.16}_{-0.12}$ & $1.31^{+0.01}_{-0.01}$ & $2.39^{+0.13}_{-0.80}$ & $1.62^{+0.02}_{-0.02}$ \\
                           & $0.83 - 1.0$  & $-0.08^{+0.01}_{-0.01}$ & $-0.08^{+0.05}_{-0.04}$ & $3.05^{+0.09}_{-0.16}$ & $-0.28^{+0.06}_{-0.04}$ & $0.23^{+0.29}_{-0.25}$ & $1.37^{+0.01}_{-0.01}$ & $1.59^{+0.05}_{-0.05}$ & $1.70^{+0.02}_{-0.02}$ \\
                           & $1.0 - 1.2$   & $-0.05^{+0.02}_{-0.02}$ & $-0.05^{+0.08}_{-0.09}$ & $2.97^{+0.09}_{-0.17}$ & $-0.26^{+0.07}_{-0.04}$ & $0.29^{+0.28}_{-0.30}$ & $1.12^{+0.03}_{-0.03}$ & $2.38^{+1.03}_{-0.47}$ & $1.29^{+0.04}_{-0.04}$ \\
                           & $1.2 - 1.65$   & $-0.01^{+0.06}_{-0.04}$ & $-0.09^{+0.52}_{-0.32}$ & $2.86^{+0.11}_{-0.22}$ & $-0.30^{+0.09}_{-0.05}$ & $0.26^{+0.29}_{-0.30}$ & $1.19^{+0.04}_{-0.05}$ & $2.41^{+0.84}_{-0.66}$ & $1.60^{+0.07}_{-0.07}$ \\
S/N$_{\rm P}$$\geq 9$       & $0.4 - 0.5$   & $-0.06^{+0.02}_{-0.04}$ & $-2.03^{+0.89}_{-0.67}$ & $2.62^{+0.05}_{-0.04}$ & $-0.25^{+0.02}_{-0.01}$ & $0.24^{+0.43}_{-0.33}$ & $1.17^{+0.02}_{-0.02}$ & $2.20^{+0.16}_{-0.79}$ & $1.95^{+0.03}_{-0.04}$ \\
                           & $0.5 - 0.65$  & $-0.18^{+0.02}_{-0.02}$ & $-0.17^{+0.10}_{-0.10}$ & $3.29^{+0.11}_{-0.17}$ & $-0.40^{+0.07}_{-0.05}$ & $0.46^{+0.31}_{-0.37}$ & $1.43^{+0.02}_{-0.02}$ & $1.60^{+0.05}_{-0.05}$ & $2.09^{+0.05}_{-0.04}$ \\
                           & $0.65 - 1.20$  & $-0.24^{+0.05}_{-0.03}$ & $-0.26^{+0.16}_{-0.19}$ & $2.99^{+0.21}_{-0.26}$ & $-0.21^{+0.10}_{-0.09}$ & $0.39^{+0.32}_{-0.30}$ & $1.15^{+0.05}_{-0.06}$ & $2.44^{+1.00}_{-0.95}$ & $1.38^{+0.07}_{-0.07}$ \\
\enddata
\tablecomments{We report all parameter estimates as the median with the 68\% confidence intervals defined by the 16th and 84th percentiles. Parameter units are the same as in Table \ref{Tab:prior}.}
\end{deluxetable*}

\begin{figure*}[htbp]
\centering
\includegraphics[width=0.6\textwidth]{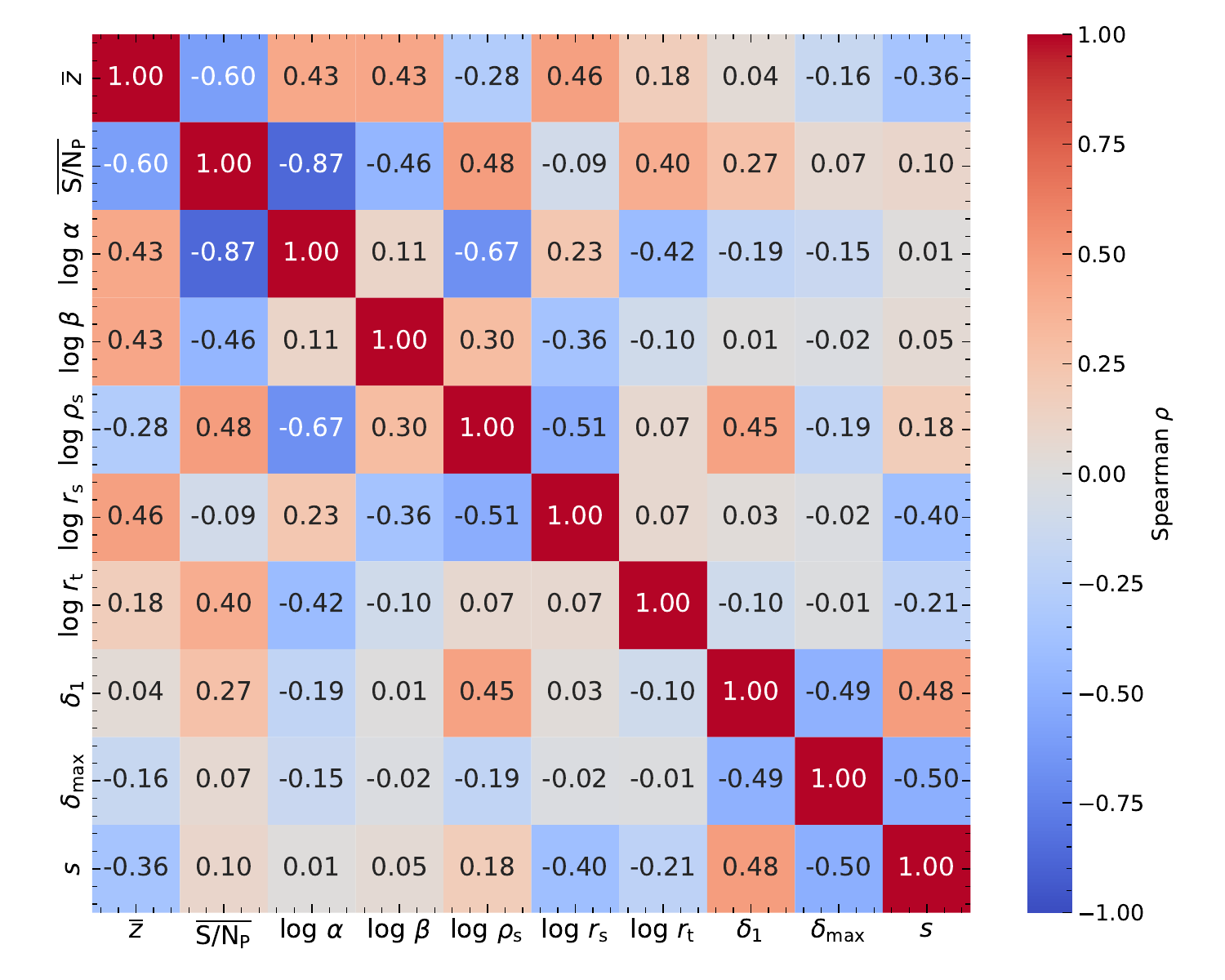}
\caption{Heatmap of Spearman's rank correlation coefficient $\rho$ from best-fit parameters in Table \ref{tab:fit_results}. }
\label{fig:spearman params}
\end{figure*}

We present the best-fit parameters for our cluster subsamples in \ref{tab:fit_results}. We explore the DK23 parameter space and its relation to the observed properties of our cluster candidates using Spearman’s rank correlation coefficients shown in Figure \ref{fig:spearman params}.

In Figure \ref{fig:spearman params}, mean signal-to-noise $\overline{\rm S/N_{\rm P}}$ shows a strong and moderate anti-correlation with the steepening parameter $\alpha$ and the truncation sharpness $\beta$, respectively, indicating a shallower and less abrupt transitions for systems with higher for clusters larger S/N$_{\rm P}$. Moderate correlations are also found between $\overline{\rm S/N_{\rm P}}$ and both $\rho_s$ and $r_t$, suggesting that higher S/N$_{\rm P}$ cluster candidates are more centrally dense and truncate at larger radius. The mean redshift $\overline{z}$ shows moderate positive correlations with $\alpha$, $\beta$, $r_s$, and moderate negative correlations with $s$, implying steeper truncation, larger central scale radius at higher redshift. For the relation between parameters themselves, $\alpha$ is strongly anti-correlated with $\rho_s$ and moderately anti-correlated with $r_t$, indicating that less steep profiles are denser and have larger truncation radius. There is also a moderate correlation between $\rho_s$ and $\delta_1$, suggesting that denser profiles are associated with higher normalization of the infalling component.

Although these correlations provide enticing insight, they are more descriptive of the observed sample. Linking the structural profiles to the underlying astrophysics of clusters requires more caution, as the observed profiles reflect a combination of selection effects, intrinsic cluster properties, and the redshift evolution of the mean density of the universe. A framework for interpreting the relationship between the DK23 parameters and cluster mass and accretion rate is presented in simulated result by \cite{2025Diemer}, who isolate structural properties of the profiles from other effects by scaling radii with $R_{\rm 200m}$ (the radius enclosing a mean density $200$ times the mean matter density of the universe). To connect the profile parameters to mass or accretion in a similar manner, MaDCoWS2 will need a precise measurement of $R_{\rm 200m}$. Thus, future work using weak-lensing analyses or simulations to obtain $R_{\rm 200m}$ will be essential for making more conclusive statements regarding the relationship between the model parameters and the underlying physical properties of our cluster sample.

Using the fitted parameters we list in Table \ref{tab:fit_results}, we show the galaxy surface densities as a function of comoving projected radii in Figure \ref{fig:dens1}, \ref{fig:dens2}, \ref{fig:dens3}, and \ref{fig:dens4} along with their best fits. We plot the comoving spherical density profiles computed from all MCMC samples. 

In Figure \ref{fig:dens1}, \ref{fig:dens2}, \ref{fig:dens3}, and \ref{fig:dens4}, the transitions between inner and outer profiles are apparent in both surface and spherical density profiles. This transition range appears at a larger radius for clusters with larger S/N$_{\rm P}$. With these clear transitions, the splashback radius can be detected as the steepest slope as shown later in \S \ref{sec:splashback radius}.

We offer a few interpretations of the density profiles displayed in Figure \ref{fig:dens1}, \ref{fig:dens2}, \ref{fig:dens3}, and \ref{fig:dens4}. 
First, the structure of the density profiles reflects the fact that cluster candidates detected by MaDCoWS2 are significant overdensities of galaxies compared to the global field. All profiles contain a core with high amplitude that gradually decreases as it extends to a larger radius. 

Second, because the density profiles are well-fit by D23, with a clearly visible break in the profile at the splashback radius, we infer that the MaDCoWS2 sample must have a sufficiently high purity over the redshift range so as to not appreciably wash out the signal. This effect is visually evident in the profiles as well as in the splashback radius determination. 

Third, the extent of the density profiles agrees with the results in \S \ref{sec:splashback radius} and in previous works that clusters with larger S/N$_{\rm P}$ are generally more extended compared to those with lower S/N$_{\rm P}$. The S/N$_{\rm P}$ scales with the galaxy overdensity at the location of the detection, and is correlated with mass in a similar way to richness at a given redshift. At fixed mass, this S/N will depend weakly on redshift due to any variation in the stellar mass threshold of the galaxy sample. This expectation is supported by the observed positive correlation of S/N$_{\rm P}$ with M$_{\rm{500c}}$ from ACT, SPT, Planck, and eRASS in \cite{2024Thongkham,2024Thongkhamb}. Further details on the relation between splashback radius and S/N$_{\rm P}$ are discussed in \S\ref{sec:splashback radius}.

Fourth, the Poisson noise associated with a given bin depends on both the number of clusters and the richness of these clusters (i.e., the number of galaxies per cluster). We can therefore improve our splashback radius analysis. The cross-correlations in this work contain small statistical uncertainties, evident in the variation in the data points. However, the typical uncertainty (5\%) is small enough that it does not impact the model fitting process. These minor fluctuations result in residuals within the expected noise level, allowing the data to be reliably fit by a two-halo theoretical density profile, such as D23.

\subsection{Effect of Uncertainty in Cluster Position and Redshift on Splashback Radius} \label{sec:extra err}

\begin{deluxetable}{cccc}[htbp!]
\tablecaption{Standard deviations of splashback radius $r_{\rm sp}$ measurements for cluster samples with scatter in position and redshift. \label{Tab:error}}
\tablehead{
\colhead{$z$} &
\colhead{$5\leq \text{S/N}_{\rm P}<7$} &
\colhead{$7\leq \text{S/N}_{\rm P}<9$} &
\colhead{$\text{S/N}_{\rm P}\geq 9$}
}
\startdata
0.4-0.5     & 0.020 & 0.044 & 0.167 \\
0.65-0.83   & 0.025 & 0.066 & \nodata \\
1.0-1.2     & 0.027 & 0.111 & \nodata \\
1.43-1.65   & 0.045 & \nodata & \nodata \\
\enddata
\tablecomments{Table is organized by redshift (first column) and S/N$_{\rm P}$ bins (remaining columns). Each entry shows the standard deviation of the three-dimensional splashback radius $r_{\rm sp}$ measurements from realizations with scatter in both position and redshift.}
\end{deluxetable}

Given the large sample sizes for the cluster and galaxy catalogs, the statistical uncertainty on the splashback radius is quite small for each bin in S/N and redshift. Nevertheless, the cross-correlations are calculated based on the assumption that the positions and redshifts of cluster detections are exact. In reality, the cluster candidates detected by MaDCoWS2 have uncertainties in both position and redshift. To assess the effect of these uncertainties on the splashback radius determination, we repeat our analysis in \S \ref{sec:measurement} for an additional 10 realizations where we include scatter in both position and redshift. The randomization process is based on the uncertainty estimation in \cite{2024Thongkhamb}. For position uncertainty, we randomly shift cluster positions using a multivariate Gaussian distribution with standard deviations of \(44^{\prime\prime}\) and \(34^{\prime\prime}\) in RA and Dec, based on the standard deviations of the offsets between MaDCoWS2 clusters and SPT clusters presented in \cite{2024Thongkham}. This is the maximum potential uncertainty from the offsets assuming all the scatter is from MaDCoWS2. For redshift uncertainty, we apply Gaussian scatter to the cluster redshifts using the standard deviation found for MaDCoWS2 photometric redshifts when compared with external spectroscopic redshifts in \citet{2024Thongkham}, $\sigma_z/(1+z)=0.027$. Due to limitations in computational time ($10$ hours per realization per subsample), we conduct this analysis only for the subsamples displayed in Table \ref{Tab:error}. We fit the relation between the uncertainty, their redshift, and S/N$_{\rm P}$ to obtain uncertainties for the remaining subsamples without reanalysis in this section.

We find that shifting cluster positions and redshifts introduces a small change in splashback radius. We report the standard deviation of this set of realizations in Table \ref{Tab:error} as a function of the S/N$_{\rm P}$ and redshift. Randomizing cluster positions flattens the radial profiles due to the fact that field galaxies contribute more. Changing the redshift of the clusters, on the other hand, keeps the clusters at the same position while the clusters are spread across redshift bins. The overdensities still stack closely on the true location, but uncertainty comes from the difference in their sizes across redshifts. Miscentered clusters could also be at incorrect redshifts. To ensure these systematics are accounted for, we thus include the uncertainties from randomization in both redshift and position in our uncertainty calculations throughout the rest of this paper. We note that these uncertainties are likely underestimated, as the model does not fully capture cluster triaxiality, filamentary environments, or correlated large-scale structure.

\subsection{Splashback Radius} \label{sec:splashback radius}

 \begin{deluxetable}{cccc}[htbp]
 \tablecaption{Splashback radii $r_{\rm sp}$ measured in this work \label{Tab:rad2}}
 \tablehead{\colhead{$z$} & \colhead{$5\leq \text{S/N}_{\rm P}<7$} & \colhead{$7\leq \text{S/N}_{\rm P}<9$} & \colhead{$\text{S/N}_{\rm P}\geq 9$}\\
 & \colhead{[$h^{-1}$ cMpc]} & \colhead{[$h^{-1}$ cMpc]} & \colhead{[$h^{-1}$ cMpc]}}
 \startdata
 0.4-0.5 &  $0.89^{+0.02(0.01)}_{-0.02(0.01)}$ & $0.93^{+0.05(0.01)}_{-0.05(0.01)}$ & $1.12^{+0.17(0.02)}_{-0.17(0.02)}$\\
 0.5-0.65 & $0.88^{+0.03(0.01)}_{-0.03(0.01)}$ & $0.86^{+0.06(0.01)}_{-0.06(0.01)}$ & $1.07^{+0.14(0.03)}_{-0.14(0.03)}$\\
 0.65-0.83 & $0.99^{+0.03(0.01)}_{-0.03(0.01)}$ & $1.17^{+0.07(0.01)}_{-0.07(0.01)}$ & \nodata \\
 0.83-1.0 & $1.08^{+0.03(0.01)}_{-0.03(0.01)}$ & $1.27^{+0.09(0.01)}_{-0.09(0.01)}$ & \nodata \\
 1.0-1.2 & $1.10^{+0.03(0.01)}_{-0.03(0.01)}$ & $1.36^{+0.11(0.02)}_{-0.11(0.02)}$ & \nodata \\
 1.2-1.43 & $1.18^{+0.04(0.01)}_{-0.04(0.01)}$ & \nodata & \nodata \\
 1.43-1.65 & $1.27^{+0.05(0.02)}_{-0.05(0.02)}$ & \nodata & \nodata \\
 \hline
 1.2–1.65 & \nodata & $1.18^{+0.12(0.04)}_{-0.12(0.04)}$ & \nodata \\
0.65–1.2 & \nodata & \nodata & $1.71^{+0.15(0.04)}_{-0.14(0.04)}$ \\
 \enddata
 \tablecomments{The three-dimensional splashback radii measurements are organized by redshift (first column) and S/N$_{\rm P}$ (three other columns). We report the combined uncertainties from jackknife resampling and those from \S \ref{sec:extra err} before parentheses. We use quadratic summation to combine the uncertainties. The uncertainties from only jacknife resampling are reported in parentheses. }
 \end{deluxetable}

 \begin{deluxetable*}{lcccc}[htbp!]
\tablecaption{Linear regression results of three-dimensional splashback radius $r_{\rm sp}$ as a function of redshift $z$ for cluster subsamples with $\text{S/N}_{\rm P}<9$, shown in both proper and comoving units.}\label{Tab:regression_twoline_split}
\tablehead{
\colhead{} & 
\multicolumn{2}{c}{$5 \leq \text{S/N}_{\rm P} < 7$} & 
\multicolumn{2}{c}{$7 \leq \text{S/N}_{\rm P} < 9$} \\
\colhead{} & 
\colhead{Proper} & \colhead{Comoving} & 
\colhead{Proper} & \colhead{Comoving}
}
\startdata
$R^2$           & $0.87$ & $0.95$ & $0.004$ & $0.64$ \\
$p$-value       & $0.002$ & $0.0002$ & $0.02$ & $0.06$ \\
$A$           & $-0.07\pm0.01$ & $0.37\pm0.04$ &  $-0.02\pm0.14$ & $0.51\pm0.19$ \\
$B$       & $0.63\pm0.01$ & $0.70\pm0.03$ & $0.65\pm0.10$ & $0.69\pm0.14$ \\
\enddata
\tablecomments{Each entry reports the linear regression statistics of $r_{\rm sp}$ vs.\ redshift $z$ for the specified S/N$_{\rm P}$ and distance convention. Slope $A$ and intercept $B$ are from weighted least-squares fits using inverse-variance weights.}
\end{deluxetable*}

\begin{deluxetable*}{cccccc}[htbp]
 \tablecaption{The splashback radius from previous works} \label{Tab:rad}
 \tablehead{\colhead{Reference} & \colhead{Sample} & \colhead{Measurement} & \colhead{Mean Redshift} & \colhead{Mean $M_{\text{\rm 200m}}$} & \colhead{$r_{\rm sp}$}\\
  & & & & \colhead{[$10^{14}\;h^{-1}$M$_\odot$]} & \colhead{[$h^{-1}$ cMpc]}}
 \startdata
 \cite{2018Chang} & DES RM & Galaxy Profile & $0.41$ & $1.69^{+0.31}_{-0.31}$ & $1.13^{+0.07}_{-0.07}$ \\
  & DES RM & Weak Lensing & $0.41$ & $1.69^{+0.31}_{-0.31}$ & $1.34^{+0.21}_{-0.21}$ \\
 \cite{2019Zurcher} & Planck SZ & Galaxy Profile & $0.18$ & $5.87^{+2.47}_{-2.87}$ & $1.78^{+0.25}_{-0.29}$ \\
 \cite{2019Contigiani} & CCCP X-ray & Weak Lensing & $0.28$ & $11.46^{+10.78}_{-6.85}$ & $2.36^{+0.74}_{-0.47}$ \\
 \cite{2019Shin} & DES RM & Galaxy Profile & $0.46$ & $5.10^{+1.56}_{-1.34}$ & $1.06^{+0.11}_{-0.09}$ \\
 & SPT SZ & Galaxy Profile & $0.49$ & $5.60^{+7.71}_{-5.99}$ & $1.24^{+0.57}_{-0.44}$ \\
  & ACT SZ & Galaxy Profile & $0.49$ & $5.10^{+4.61}_{-4.43}$ & $1.32^{+0.40}_{-0.38}$ \\
 \cite{2020Murata} & HSC CAMIRA & Galaxy Profile & $0.57$ & $1.71^{+0.08}_{-0.07}$ & $1.50^{+0.19}_{-0.18}$ \\
 \cite{2021Shin} & ACT SZ & Weak Lensing & $0.46$ & $4.62^{+3.50}_{-4.90}$ & $1.19^{+0.30}_{-0.42}$ \\
 & ACT SZ & Galaxy Profile & $0.46$ & $4.62^{+1.10}_{-2.40}$ & $1.13^{+0.09}_{-0.13}$ \\
 \cite{2023Rana} & eFEDS X-ray & Galaxy Profile & $<0.75$ & $14.52^{+0.06}_{-0.06}$ & $1.45^{+0.30}_{-0.26}$ \\
 \cite{2024Xu} & ZOU21z2m1 & Weak Lensing & $0.5$ & $0.71^{+0.15}_{-0.15}$ & $0.60^{+0.25}_{-0.25}$ \\
 & ZOU21z2m2 & Weak Lensing & $0.51$ & $1.62^{+0.21}_{-0.21}$ & $0.97^{+0.45}_{-0.45}$ \\
 & YANG21z2m1 & Weak Lensing & $0.47$ & $0.37^{+0.16}_{-0.16}$ & $0.51^{+0.26}_{-0.26}$ \\
 & YANG21z2m2 & Weak Lensing & $0.47$ & $0.60^{+0.19}_{-0.19}$ & $0.70^{+0.39}_{-0.39}$ \\
 \enddata
 \tablecomments{Results from previous works were standardized as in \cite{2024Xu}. We include only a subset of results from \cite{2024Xu} that explore the mass and redshift ranges similar to our work. The splashback radius is three-dimensional.}
 \end{deluxetable*}
 
\begin{figure*}[htbp]
\centering
\includegraphics[width=1.5\columnwidth]{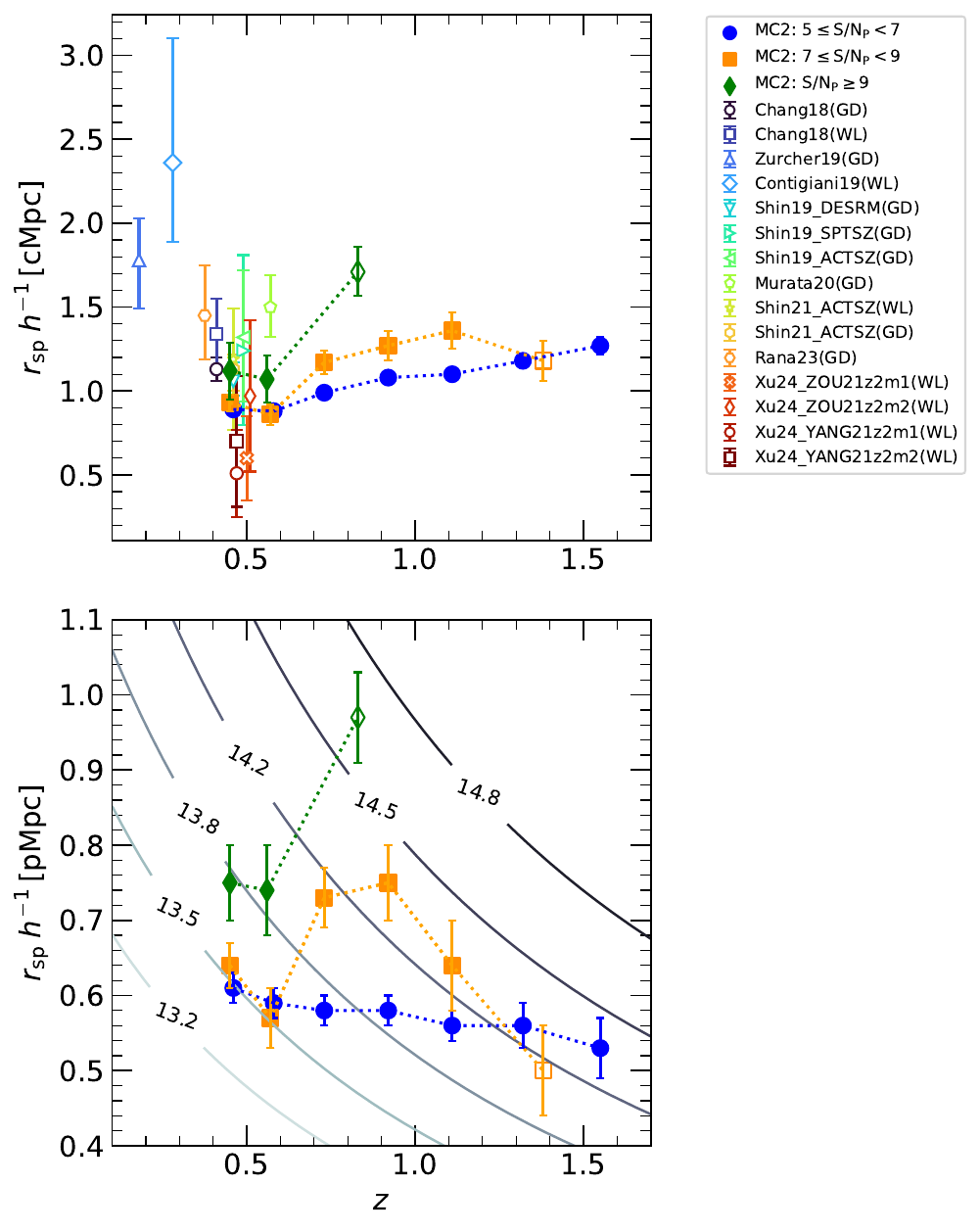}
\caption{Three-dimensional splashback radius $r_{\rm sp}$ as a function of redshift. Solid blue circles, orange squares, and green diamonds indicates $5\leq \text{S/N}_{\rm P}<7$, $7\leq \text{S/N}_{\rm P}<9$, and $\text{S/N}_{\rm P}\geq9$. Open orange square and green diamond indicate binning schemes that cover remaining combined ranges for high-$\text{S/N}_{\rm P}$ subsamples. Errors are estimated from the quadratic summation of uncertainties from jackknife resampling and uncertainties estimated in \S \ref{sec:extra err}. The top panel shows splashback radii determined in comoving frames similar to previous works in the literature in Table \ref{Tab:rad} (open markers). The bottom panel shows the radius when determined in the proper frame. The contours in the bottom panel show $M_{\rm 200m}$ derived using \cite{2015More} splashback model in units of $\log(h^{-1}M_\odot)$.}
\label{fig:rsp_vs_z}
\end{figure*}

We follow the description of the splashback radius introduced in \cite{2015More} which defines the splashback radius as the location where the cluster density profile is the steepest. We calculate the logarithmic derivatives of the density profiles and show them in Figure \ref{fig:dens1}, \ref{fig:dens2}, \ref{fig:dens3}, and \ref{fig:dens4}. We show our splashback radius estimates as lines of the same colors in these Figures. A strong dip, always extending to $\mathrm{d}\;\log\;\rho_{g}\;/\;\mathrm{d}\;\log\;r < -3$, is observed in all subsamples in Figure \ref{fig:dens1}, \ref{fig:dens2}, \ref{fig:dens3}, and \ref{fig:dens4}, indicating a clear signal from the splashback radius \citep{2017Baxter}. We show our splashback measurements in Table \ref{Tab:rad2}, and plot the splashback radius as determined in the comoving and proper frames as a function of redshift in Figure \ref{fig:rsp_vs_z}.

We test for a significant trend in splashback radii as a function of redshift by performing a weighted least-squares (WLS) regression using the Python package \texttt{statsmodels} \citep{seabold2010statsmodels}, focusing on the subsamples with $\text{S/N}_{\rm P}<9$. The results of this analysis are presented in Table \ref{Tab:regression_twoline_split}. We find that we can reject a flat model with $95\%$ confidence (p-values below $0.05$ for almost all analysis configurations). The splashback radius in the comoving frame increases with redshift, which suggests a positive correlation between $r_{\rm sp}$ and redshift in comoving coordinate system. Increases in $r_{\rm sp}$ for higher $\text{S/N}_{\rm P}$ are also observed in almost all redshift bins, especially when comparing $\text{S/N}_{\rm P}\geq9$ to lower $\text{S/N}_{\rm P}$ subsamples. In physical units (or proper frame), the splashback radius decreases with redshift while still exhibiting a positive correlation with $\text{S/N}_{\rm P}$. This is supported by the fact that we can also reject a flat model with $95\%$ confidence for our results in the proper frame. The difference in the redshift trends of the comoving and proper sizes arises from the $(1+z)$ factor, which accounts for the change in physical scale due to cosmic expansion.

The top panel of Figure \ref{fig:rsp_vs_z} highlights the redshift reach of our analysis compared to previous results in the literature from \cite{2018Chang,2019Zurcher,2019Contigiani,2019Shin,2020Murata,2021Shin,2023Rana,2024Xu}, also presented in Table \ref{Tab:rad}. With MaDCoWS2, we extend splashback radius measurements based on galaxy density beyond $z > 1$, achieving competitive statistical uncertainties.

Nonetheless, we emphasize that the results shown in Table \ref{Tab:rad2} and the top panel of Figure \ref{fig:rsp_vs_z} are specific to the MaDCoWS2 sample and may be affected by selection effects already mentioned in \ref{sec:dens prof}, which would benefit from further investigations using weak lensing or simulations.

\subsection{Discussion} \label{sub:discussion}

While the apparent increase of the physical splashback radius (calculated in the proper frame) with decreasing redshift could suggest an evolutionary trend towards more massive clusters that is expected from the evolution of a halo mass function \citep{2008Tinker,2016Bocquet,2016Despali}, the splashback radius depends intricately on peak height ($\nu$), mass, and redshift \citep{2015More,2025Diemer}, precluding a straightforward interpretation. To better interpret the relationship between splashback radius, S/N$_{\rm P}$, and cluster redshift, we adopt the empirical physical splashback radius–mass relation of \cite{2015More}, use it to infer cluster masses, and compare them with estimates from \cite{2024Thongkhamb} and weak-lensing results from the literature (Table \ref{Tab:rad}).

In \cite{2015More}, the three-dimensional splashback radius $r_{\rm sp}$ and mass can be calculated by
\begin{align}
\label{eq:omegaR2}
r_{\rm sp} &= 0.81 \left(1 + 0.97 e^{\nu / 2.44} \right) r_{\rm 200m}\\
M_{ sp} &= 0.82 \left(1 + 0.63 e^{\nu / 3.52} \right) M_{\rm 200m}
\end{align}
where $\nu$ is the peak height which is defined as $\delta_c/\sigma(M_{\rm 200m},z)$, and $r_{\rm 200m}(z)$ is the radius enclosing a mean density $200$ times the mean matter density of the universe. We convert the proper splashback radius into $M_{\rm 200m}$ using the \texttt{splashback.splashbackRadius} function of the \texttt{COLOSSUS} \citep{2018Diemer} package. The function applies the splashback model from \cite{2015More} to estimate splashback radius from mass/accretion rate and redshift. The model from \cite{2015More} defines the splashback radius as the steepest slope of the density profile, and is calibrated using averaged halo profiles from $\Lambda$CDM cosmological simulations. The simulations are run using GADGET2 \citep{2005Springel} described in \cite{2014Diemer}. For each redshift, we construct a grid of $M_{\rm 200m}$ and $r_{\rm sp}$ and estimate $M_{\rm 200m}$ from a given $r_{\rm sp}$. We show the mass grid as contour lines in the lower panel of Figure \ref{fig:rsp_vs_z}. We present the average $M_{\rm 200m}$ in a given MaDCoWS2 S/N$_{\rm P}$ bin as a function of cluster redshift in Table \ref{Tab:m200}. The uncertainties of $M_{\rm 200m}$ are estimated from the 1$\sigma$ upper and lower errors of the three-dimensional splashback radii in physical units.

 \begin{deluxetable}{cccc}[htbp]
 \tablecaption{$M_{\rm 200m}$ measured from three-dimensional splashback radius $r_{\rm sp}$ in this work using splashback model from \cite{2015More}} \label{Tab:m200}
 \tablehead{\colhead{$z$} & \colhead{$5\leq \text{S/N}_{\rm P}<7$} & \colhead{$7\leq \text{S/N}_{\rm P}<9$} & \colhead{$\text{S/N}_{\rm P}\geq 9$}}
 \startdata
 0.4-0.5 &  $13.49^{+0.05}_{-0.05}$ & $13.55^{+0.07}_{-0.07}$ & $13.79^{+0.10}_{-0.11}$\\
 0.5-0.65 &  $13.58^{+0.05}_{-0.05}$ & $13.52^{+0.10}_{-0.11}$ & $13.91^{+0.12}_{-0.13}$\\
 0.65-0.83 & $13.72^{+0.05}_{-0.05}$ & $14.08^{+0.08}_{-0.09}$ & \nodata \\
 0.83-1.0 & $13.92^{+0.06}_{-0.06}$ & $14.33^{+0.1}_{-0.11}$ & \nodata \\
 1.0-1.2 & $14.05^{+0.06}_{-0.06}$ & $14.26^{+0.14}_{-0.16}$ & \nodata \\
 1.2-1.43 & $14.24^{+0.08}_{-0.09}$ & \nodata & \nodata \\
 1.43-1.65 & $14.33^{+0.11}_{-0.12}$ & \nodata & \nodata \\
 \hline
 1.2–1.65 & \nodata & $14.11^{+0.18}_{-0.20}$ & \nodata \\
0.65–1.2 & \nodata & \nodata & $14.65^{+0.10}_{-0.10}$ \\
 \enddata
 \tablecomments{Similar to Table \ref{Tab:cluster sample}, we organize $M_{\rm 200m}$ by redshift (first column) and S/N$_{\rm P}$ (three other columns). $M_{\rm 200m}$ is in the units of $\log\;(h^{-1}M_{\odot})$. }
 \end{deluxetable}

\subsubsection{Comparison with the MaDCoWS2 DR2 Mass–S/N$_{\rm P}$ Relation} \label{sub:compare to DR2}

Considering the positive relation between MaDCoWS2 S/N$_{\rm P}$ and $M_{\rm 500c}$ estimated from surveys such as ACT \citep{2021Hilton}, SPT \citep{2015Bleem,2019Bocquet,2020Bleem,2024Bleem}, and eRASS \citep{2024Bulbul,2024Kluge} observed in \S $5.4$ of \cite{2024Thongkhamb}, it is expected that subsamples with higher $\text{S/N}_{\rm P}$ in the same redshift range have larger $M_{\rm 200m}$ based on the \citet{2015More} relation. It is logical that clusters with higher S/N$_{\rm P}$, which serves as a proxy for the number density of galaxies in the area, would be more massive. 

Additionally, the increase of $M_{\rm 200m}$, derived from \cite{2015More} model, with $\overline{z}$ is broadly similar to the fit results reported in \citet{2024Thongkham,2024Thongkhamb}. $M_{\rm 500c}$, derived from ACT, PSZ2 \citep{2016Planck}, and eRASS, shows a positive power-law correlation with $1+z$ when fitted as a function of S/N$_{\rm P}$. In most relations we investigated in \cite{2024Thongkhamb}, we found an overall increase of cluster $M_{\rm 500c}$, derived from MaDCoWS2 S/N$_{\rm P}$ and $M_{\rm 500c}$ relation, with redshift, even though the slopes and intercepts differ from this work depending on the comparison sample. This apparent increase of $M_{\rm 200m}$ and $M_{\rm 500c}$ with redshift may partly reflect selection effects connected to photometric catalog limits. Since MaDCoWS2 requires optical detections in addition to CatWISE, the effective luminosity threshold rises with redshift. At the highest redshifts, only the most $z$-band luminous galaxies are detected, possibly biasing the sample toward more massive clusters that host them \citep{2017Zhang,2017Mulroy}.

Although the agreement on the direction of the trends is encouraging, a few caveats are worth noting. First, because of scatter in our mass proxy (S/N$_{\rm P}$), the inferred masses are similar for the $5\leq \text{S/N}_{\rm P}<7$ and $7\leq \text{S/N}_{\rm P}<9$ bins. Second, we note that $M_{\rm 200m}$ values derived based upon the \cite{2015More} relation are systematically lower than those inferred from ACT, SPT, and eRASS calibrations in \cite{2024Thongkham,2024Thongkhamb}. We further explore this offset in the next section.

\subsubsection{Comparison with results in literature that include weak-lensing masses} \label{sub:compare to literature}

Compared to the comoving splashback radii reported in previous observational studies, as summarized in Table \ref{Tab:rad}, the comoving splashback radii estimated for MaDCoWS2 clusters at $z<0.75$ are in the middle of the observed mass range. Earlier studies prior to \citet{2024Xu} usually focused on more massive clusters ($M_{\rm 200m} \gtrsim 10^{14} \;h^{-1},\mathrm{M}_\odot$) to ensure robust statistical analyses, resulting in splashback radius measurements exceeding $1 \;h^{-1}$ cMpc. On the other hand, the cluster catalogs of \citet{2024Xu}, constructed from Legacy Surveys and WISE photometric data \citep{2021Zou,2021Yang}, are more comparable to MaDCoWS2 in terms of selection and mass range, though a few of their samples include clusters at a lower mass range. 

In the context of the comoving splashback radius–$M_{\rm 200m}$ relation, our finding that larger comoving splashback radii occur with increasing S/N$_{\rm P}$ or redshift implies correspondingly larger $M_{\rm 200m}$. This is consistent with the evolution of $M_{\rm 200m}$ and $M_{\rm 500c}$ with redshift and S/N$_{\rm P}$, based on the calculation using relations from \citet{2015More} and \citet{2024Thongkhamb}, respectively. However, deriving $M_{\rm 200m}$ for MaDCoWS2 using the relation from \citet{2015More} also results in some discrepancies. At fixed S/N$_{\rm P}$, these MaDCoWS2 $M_{\rm 200m}$ estimates based on the \citet{2015More} relation, especially at $z<1$, are systematically lower than the expectation from the comoving splashback radius and the weak-lensing $M_{\rm 200m}$ in Table \ref{Tab:rad}, as well as our other mass estimates as discussed in the previous section. We fit a power-law model, 
\begin{equation}
\label{eq:M200m}
M_{\rm 200m} = a r_{\rm sp}^b
\end{equation}
to the observed weak-lensing–splashback radius data in Table \ref{Tab:rad}, finding $a = 1.59\pm0.36$ and $b = 3.55\pm0.50$. Using this relation, the $M_{\rm 200m}$ values inferred from our splashback radii are, on average, $2.4$, $2.6$, and $2.8$ times higher than those from the \cite{2015More} relation for $5\leq$ S/N$_{\rm P}$ $<7$, $7\leq$ S/N$_{\rm P}$ $<9$, and S/N$_{\rm P}$ $\geq9$, respectively. A similar discrepancy was seen in the observed mass ratio–peak height relation of \citet{2024Xu}, suggesting that the \citet{2015More} relation may underestimate $M_{\rm 200m}$ in some samples.

Several explanations can be explored for this discrepancy. Our galaxy selection function might select galaxy populations that bias the splashback radii to smaller values \citep{2021Adhikari,2022O'Neil}. A related argument is that baryonic effects, such as dynamical friction, may also contribute to the divergence between our results and the \cite{2015More} relation in the regime probed by our observational data \citep{2020Xhakaj,2025O'Shea}. The most plausible explanation likely involves a combination of these effects, together with systematic uncertainties in the observations and assumptions inherent in the simulations.

Our observed evolution of the splashback radius is likely due to a combination of factors rather than purely arising from physical evolution. An improved weak lensing mass calibration and characterization of redshift-dependent selection effects will be important to facilitate the use of the splashback radius as a tool for determining $M_{\rm 200m}$. Such a calibration could be accomplished with deep, wide-field data from the $\textit{Euclid}$ or $\textit{Nancy Grace Roman Space Telescope}$ ($\textit{Roman}$) as they become available.

\subsection{Future improvement}

One potential improvement to the analysis conducted here would be a direct measurement of cluster mass based on a profile that includes all matter. With mass and spherical overdensity radius estimated from weak lensing (i.e., $r_{\rm 200m}$), the splashback radius delivered in this work could be put in a better context. This would improve knowledge of the ratio of splashback radius and weak lensing radius relative to peak height and accretion rate and better validate the relation from cosmological simulations such as \cite{2015More,2025Diemer}. We could also combine the splashback radius and the shear profile within the radius to calibrate the splashback radius-lensing mass relation. 

These improvements could be achieved using shape measurements from the Dark Energy Survey Year 3 \citep{2021Gatti}, Kilo Degree Survey \citep{2021Giblin}, and the Hyper Supreme Camera wide survey \citep{2022Li}, for the redshift and mass range over which weak lensing analyses are possible with these data sets. Further benefits are expected from next-generation facilities such as $\textit{Euclid}$, the Rubin Observatory, and $\textit{Roman}$, which will provide higher-precision shape measurements.

\section{Conclusions} \label{sec:conclusion}

We have investigated the galaxy distribution and splashback feature in the MaDCoWS2 cluster sample. Surface density profiles based on cross-correlations of the galaxies from CatWISE2020 and clusters from MaDCoWS2 were measured in different redshift and S/N$_{\rm P}$ bins. We fit the surface density to a model with orbiting and infalling terms using Bayesian analysis. Using the MCMC samples, we obtained spherical density profiles and determined the splashback radii from their steepest slopes. We summarize our main findings as follows:

\begin{itemize}
    \item We detect splashback radii out to z=1.65, which is the farthest that such a feature has been found. Given the large sample sizes, we are able to derive splashback radii with $2\%$ statistical uncertainties in redshift bins spanning the range $0.4-1.65$ and in multiple S/N$_{\rm P}$ bins at the low redshift end of this range.
    
    \item We confirm that the measured splashback radius increases with S/N$_{\rm P}$, consistent with the positive correlation between S/N$_{\rm P}$ and cluster mass indicated in \cite{2024Thongkham,2024Thongkhamb}.
    
    \item We investigate the redshift dependence of the splashback radius, finding that in physical units it decreases by about $13\%$ from $z=0.4$ to $z=1.65$ for the lowest S/N$_{\rm P}$ bin. In comoving units, it increases by $43\%$.
    
    \item Given the redshift dependence of the splashback radius, if we use the scaling relation of \citet{2015More} then this implies that the $M_{\rm 200m}$ threshold of MaDCoWS2 approximately increases by a factor of $5$ from $\overline{z}=0.46$ to $1.55$. A similar positive mass-redshift trend was found in our previous works \citep{2024Thongkham,2024Thongkhamb} although with larger mass. Nevertheless, we note that a more detailed investigation of the selection function from simulations or weak lensing is needed to more robustly explain the evolution of splashback radius and cluster physical properties.
    
    \item The redshift dependence of $M_{\rm 200m}$ derived here using the \citet{2015More} relation has a similar redshift trend to that which was found in \citet{2024Thongkham,2024Thongkhamb} based on literature masses for SZ and X-ray clusters in MaDCoWS2. The $M_{\rm 200m}$ computed using the \citet{2015More} relation are however systematically lower than the prediction based on the MaDCoWS2 scaling relations. Additionally, observational calibrations of the $r_{\rm sp}-M_{\rm 200m}$ scaling relation using lensing masses indicate that, for a given $r_{\rm sp}$, the \citet{2015More} relation may be underpredicting the masses. A weak-lensing calibration with more robust mass measurements will be crucial to understand this discrepancy.
\end{itemize}

The analysis in this work shows the statistical power of a large sample of galaxy clusters detected by an algorithm such as PZWav, using the full photometric redshift information, in constraining splashback radii. Nevertheless, future analyses with direct mass measurements are still needed to robustly use splashback radii as a tool for cluster mass estimation. With $\textit{Euclid}$ and Rubin observing a large area of the sky to great depth at high resolution, a better understanding of the halos of galaxy clusters up to high redshift can be achieved in the near future.

\begin{acknowledgments}
This work was supported by the Science and Technology Development Fund of Thailand Science Research and Innovation (TSRI), through the National Astronomical Research Institute of Thailand (Public Organization), Fiscal Year 2025 (ST68), under the project “Enhancing Thailand’s Space Situation through Astronomical Technology” and by the Fundamental Fund of Thailand Science Research and Innovation (TSRI) through the National Astronomical Research Institute of Thailand (Public Organization), Fiscal Year 2026 (FF69). This research was supported by the Korea Astronomy and Space Science Institute under the R$\&$D program (Project No. $2026-1-831-02$) supervised by the Korea AeroSpace Administration. This material is based on work supported by the National Science Foundation under Grant No. 2008367. The work of P.E. and D.S. was carried out at the Jet Propulsion Laboratory, California Institute of Technology, under a contract with the National Aeronautics and Space Administration (80NM0018D0004). The authors acknowledge University of Florida Research Computing for providing computational resources and support that have contributed to the research results reported in this publication.

This publication makes use of data products from the \textit{Wide-field Infrared Survey Explorer}, which is a joint project of the University of California, Los Angeles, and the Jet Propulsion Laboratory/California Institute of Technology, funded by the National Aeronautics and Space Administration.

This research has made use of the NASA/IPAC Infrared Science Archive, which is funded by the National Aeronautics and Space Administration and operated by the California Institute of Technology.

This research is based upon the Dark Energy Camera Legacy Survey (DECaLS; PIs: David Schlegel and Arjun Dey). The Legacy Surveys imaging of the DESI footprint is supported by the Director, Office of Science, Office of High Energy Physics of the U.S. Department of Energy under Contract No. DE-AC02-05CH1123, by the National Energy Research Scientific Computing Center, a DOE Office of Science User Facility under the same contract; and by the U.S. National Science Foundation, Division of Astronomical Sciences under Contract No.AST-0950952 to NOAO.

This project used data obtained with the Dark Energy Camera (DECam), which was constructed by the Dark Energy Survey (DES) collaboration. Funding for the DES Projects has been provided by the U.S. Department of Energy, the U.S. National Science Foundation, the Ministry of Science and Education of Spain, the Science and Technology Facilities Council of the United Kingdom, the Higher Education Funding Council for England, the National Center for Supercomputing Applications at the University of Illinois at Urbana-Champaign, the Kavli Institute of Cosmological Physics at the University of Chicago, Center for Cosmology and Astro-Particle Physics at the Ohio State University, the Mitchell Institute for Fundamental Physics and Astronomy at Texas A$\&$M University, Financiadora de Estudos e Projetos, Fundação Carlos Chagas Filho de Amparo, Financiadora de Estudos e Projetos, Fundação Carlos Chagas Filho de Amparo à Pesquisa do Estado do Rio de Janeiro, Conselho Nacional de Desenvolvimento Científico e Tecnológico and the Ministério da Ciência, Tecnologia e Inovação, the Deutsche Forschungsgemeinschaft and the Collaborating Institutions in the Dark Energy Survey. The Collaborating Institutions are Argonne National Laboratory, the University of California at Santa Cruz, the University of Cambridge, Centro de Investigaciones Enérgeticas, Medioambientales y Tecnológicas–Madrid, the University of Chicago, University College London, the DES-Brazil Consortium, the University of Edinburgh, the Eidgenössische Technische Hochschule (ETH) Zürich, Fermi National Accelerator Laboratory, the University of Illinois at Urbana-Champaign, the Institut de Ciències de l'Espai (IEEC/CSIC), the Institut de Física d'Altes Energies, Lawrence Berkeley National Laboratory, the Ludwig-Maximilians Universität München and the associated Excellence Cluster Universe, the University of Michigan, the National Optical Astronomy Observatory, the University of Nottingham, the Ohio State University, the University of Pennsylvania, the University of Portsmouth, SLAC National Accelerator Laboratory, Stanford University, the University of Sussex, and Texas A$\&$M University.

The Photometric Redshifts for the Legacy Surveys (PRLS) catalog used in this paper was produced thanks to funding from the U.S. Department of Energy Office of Science, Office of High Energy Physics via grant DE-SC0007914.

The Siena Galaxy Atlas was made possible by funding support from the U.S. Department of Energy, Office of Science, Office of High Energy Physics under Award Number DE-SC0020086 and from the National Science Foundation under grant AST-1616414.

\end{acknowledgments}
\vspace{5mm}

\software{PZWav, Astropy \citep{astropy:2013, astropy:2018, astropy:2022}, Matplotlib \citep{Hunter:2007}, NumPy \citep{harris2020array}, pandas \citep{the_pandas_development_team_2023_8092754,mckinney-proc-scipy-2010}, lmfit \citep{newville_matthew_2014_11813}, SciPy \citep{2020Virtanen}, treecorr \citep{2004Jarvis,2015Jarvis}, emcee \citep{2013Foreman}, COLOSSUS \citep{2018Diemer}, statsmodel \citep{seabold2010statsmodels}}

\newpage
\clearpage

\appendix
\section{Projected Splashback Radius from Galaxy Surface Density in Proper Frame} \label{sec:bright stars}

In addition to the splashback radii presented in \S\ref{sec:splashback radius}, we also provide measurements derived from the galaxy surface density, $\Sigma_g$, in proper frame. The splashback radius is identified as the location of the steepest slope in the best-fit galaxy surface density profile. These measurements are shown in Table \ref{Tab:proprad2d}. They serve as a basis for comparison with the radii determined from surface brightness profiles presented in \cite{2026Trudeau}.

 \begin{deluxetable}{cccc}[htbp]
 \tablecaption{Splashback radii $R_{\rm sp}$ measured directly from surface densities in proper frame\label{Tab:proprad2d}}
 \tablehead{\colhead{$z$} & \colhead{$5\leq \text{S/N}_{\rm P}<7$} & \colhead{$7\leq \text{S/N}_{\rm P}<9$} & \colhead{$\text{S/N}_{\rm P}\geq 9$}\\
 & \colhead{[$h^{-1}$ cMpc]} & \colhead{[$h^{-1}$ cMpc]} & \colhead{[$h^{-1}$ cMpc]}}
 \startdata
 0.4-0.5 &  $0.45^{+0.01}_{-0.01}$ & $0.48^{+0.02}_{-0.02}$ & $0.58^{+0.3}_{-0.03}$\\
 0.5-0.65 & $0.46^{+0.01}_{-0.01}$ & $0.45^{+0.02}_{-0.02}$ & $0.55^{+0.03}_{-0.03}$\\
 0.65-0.83 & $0.44^{+0.01}_{-0.01}$ & $0.51^{+0.02}_{-0.02}$ & \nodata \\
 0.83-1.0 & $0.45^{+0.01}_{-0.01}$ & $0.49^{+0.02}_{-0.02}$ & \nodata \\
 1.0-1.2 & $0.42^{+0.01}_{-0.01}$ & $0.46^{+0.03}_{-0.03}$ & \nodata \\
 1.2-1.43 & $0.39^{+0.01}_{-0.01}$ & \nodata & \nodata \\
 1.43-1.65 & $0.40^{+0.02}_{-0.02}$ & \nodata & \nodata \\
 \hline
 1.2–1.65 & \nodata & $0.39^{+0.02}_{-0.02}$ & \nodata \\
0.65–1.2 & \nodata & \nodata & $0.67^{+0.03}_{-0.03}$ \\
 \enddata
 \tablecomments{Similar to Table \ref{Tab:rad2}. The splashback radii measurements are organized by redshift (first column) and S/N$_{\rm P}$ (three other columns). We report the combined uncertainties from jackknife resampling and those from \S \ref{sec:extra err}.}
 \end{deluxetable}

\newpage
\clearpage

\bibliography{MaDCoWS2splashback}{}
\bibliographystyle{aasjournalv7}

\end{document}